\documentclass[twocolumn,times,twocolappendix]{aastex631}

\usepackage{amsmath}

\usepackage[]{newtxtext}
\usepackage[]{newtxmath}

\usepackage{boogaard-commands}

\defcitealias{Tan2014}{T14}
\defcitealias{Cortzen2020}{C20}
\defcitealias{Carilli2010}{C10}
\defcitealias{Carilli2011}{C11}
\defcitealias{Hodge2012}{H12}
\defcitealias{Hodge2015}{H15}
\defcitealias{Magdis2011}{M11}
\defcitealias{Pope2006}{P06}
\defcitealias{Perera2008}{P08}
\defcitealias{Kolupuri2025}{K25}

\received{August 5, 2025}
\revised{October 7, 2025}
\accepted{October 17, 2025}

\shorttitle{}
\shortauthors{Boogaard et al.}
\graphicspath{{./}{figures/}}

\begin{document}

\title{Resolving the dusty star-forming galaxy GN20 at $z=4.055$ with
  NOEMA and JWST: A similar distribution of stars, gas and dust
  despite distinct apparent profiles}

\newcommand{\MPIA}{\affiliation{Max Planck Institute for Astronomy, K\"onigstuhl 17,  69117 Heidelberg, Germany}}

\newcommand{\DAWN}{\affiliation{Cosmic Dawn Center (DAWN), Denmark}}

\newcommand{\DTU}{\affiliation{DTU Space, Technical University of Denmark, Elektrovej, Building 328, 2800, Kgs. Lyngby, Denmark}}

\newcommand{\DARK}{\affil{DARK, Niels Bohr Institute, University of Copenhagen, Jagtvej 128, 2200 Copenhagen, Denmark}}

\newcommand{\NBI}{\affil{Niels Bohr Institute, University of Copenhagen, Jagtvej 128, 2200 Copenhagen, Denmark}}

\newcommand{\Stockholm}{\affiliation{Department of Astronomy, Stockholm University, Oscar Klein Centre, AlbaNova University Centre, 106 91 Stockholm, Sweden}}

\newcommand{\NRAO}{\affiliation{National Radio Astronomy Observatory, Pete V. Domenici Array Science Center, P.O. Box O, Socorro, NM 87801, USA}}

\newcommand{\CABcda}{\affiliation{Centro de Astrobiolog\'{i}a (CAB), CSIC-INTA, Ctra. de Ajalvir km 4, Torrej\'{o}n de Ardoz, E-28850, Madrid, Spain}}

\newcommand{\CABcbdc}{\affiliation{Centro de Astrobiolog\'{i}a (CAB), CSIC-INTA, Camino Bajo del Castillo s/n, 28692 Villanueva de la Ca\~{n}ada, Madrid, Spain}}

\newcommand{\ESAC}{\affiliation{Telespazio UK for the European Space Agency, ESAC, Camino Bajo del Castillo s/n, 28692 Villanueva de la Ca\~nada, Spain}}

\newcommand{\Kapteyn}{\affiliation{Kapteyn Astronomical Institute, University of Groningen, P.O. Box 800, 9700AV Groningen, The Netherlands}}

\newcommand{\Leiden}{\affil{Leiden Observatory,  Leiden University, PO Box 9513, NL-2300 RA Leiden, The Netherlands}}

\newcommand{\UCL}{\affiliation{Dept. of Physics and Astronomy, University College London, Gower Street, London WC1E 6BT, United Kingdom}}

\newcommand{\Edinburgh}{\affiliation{UK Astronomy Technology Centre, Royal Observatory Edinburgh, Blackford Hill, Edinburgh EH9 3HJ, UK}}

\newcommand{\ITA}{\affil{Institute for Theoretical Physics, Heidelberg University, Philosophenweg 12, D–69120, Heidelberg, Germany}}

\newcommand{\Cologne}{\affil{I.Physikalisches Institut der Universit\"{a}t zu K\"{o}ln, Z\"{u}lpicher Str. 77, 50937 K\"{o}ln, Germany}}

\newcommand{\MPIfR}{\affil{Max Planck Institute for Radiosastronomie, Auf dem H\"{u}gel 69, 53121 Bonn, Germany}}

\newcommand{\Geneva}{\affil{Departement d'Astronomie, University of Geneva, Chemin Pegasi 51, 1290 Versoix, Switzerland}}

\newcommand{\Marseille}{\affil{Aix Marseille Univ, CNRS, CNES, LAM, Marseille, France}}

\newcommand{\SPL}{\affil{School of Physics \& Astronomy, Space Research Centre, Space Park Leicester, University of Leicester, 92 Corporation Road, Leicester LE4 5SP, UK}}

\newcommand{\Dublin}{\affil{Dublin Institute for Advanced Studies, Astronomy \& Astrophysics Section, 31 Fitzwilliam Place, Dublin 2, Ireland}}

\newcommand{\ParisDaddi}{\affil{CEA, IRFU, DAp, AIM, Universit\'e Paris-Saclay, Universit\'e Paris Cit\'e, Sorbonne Paris Cit\'e, CNRS, 91191 Gif-sur-Yvette, France}}

\newcommand{\STScI}{\affil{Space Telescope Science Institute (STScI), 3700 San Martin Drive, Baltimore, MD 21218, USA}}
 
\correspondingauthor{Leindert Boogaard}
\email{boogaard@strw.leidenuniv.nl}

\author[0000-0002-3952-8588]{Leindert A. Boogaard} \Leiden

\author[0000-0003-4793-7880]{Fabian Walter} \MPIA \NRAO

\author[0000-0003-4678-3939]{Axel Wei\ss} \MPIfR

\author[0000-0002-9090-4227]{Luis Colina} \CABcda

\author[0000-0001-6586-8845]{Jacqueline Hodge} \Leiden

\author[0000-0001-8068-0891]{Arjan Bik} \Stockholm

\author[0000-0003-2119-277X]{Alejandro Crespo G\'{o}mez} \STScI

\author[0000-0002-3331-9590]{Emanuele Daddi} \ParisDaddi

\author[0000-0002-4872-2294]{Georgios E. Magdis} \DAWN \DTU \NBI

\author[0000-0001-5492-4522]{Romain A. Meyer} \Geneva

\author[0000-0002-3005-1349]{G\"{o}ran \"{O}stlin} \Stockholm

\begin{abstract}
  We present high-resolution (0\farcs13--0\farcs23) NOEMA observations
  of the dust continuum emission at 1.1\,mm (rest-frame 220\,\micron)
  and JWST/NIRCam and MIRI imaging of the $z=4.055$ starburst galaxy
  GN20.  The sensitive NOEMA imaging at 1.6\,kpc resolution reveals
  extended dust emission, $\approx$14\,kpc in diameter
  ($r_e\approx2.5$\,kpc, $b/a=0.5$), that is centrally asymmetric
  and clumpy.  The dust emission is as extended as the stellar
  emission and the molecular gas traced by $^{12}$CO(2--1), with a
  common center, and is brightest in the strongly-obscured nuclear
  part of the galaxy.  Approximately one-third of the total dust
  emission emerges from the nucleus and the most prominent clump to
  the south, and (only) 60\% from the central $3.5\times1.5$\,kpc
  (0\farcs5--0\farcs2), implying that the starburst is very extended.
  The combined JWST and NOEMA morphology suggests GN20 experienced a
  recent interaction or merger, likely invigorating the starburst.
  The radial surface brightness profiles of the molecular gas and
  near-infrared stellar emission are similar, while in contrast, the
  dust emission appears significantly more concentrated.  Through
  self-consistent radiative transfer modeling of the integrated and
  resolved $^{12}$CO and dust emission, we derive an
  $\Mmol=2.9^{+0.4}_{-0.3}\times10^{11}$\,\Msun\ with
  $\aco=2.8^{+0.5}_{-0.3}$.  We find the extended dust implies a lower
  global dust optical depth than previously reported, but a high dust
  mass of $\Mdust=5.7^{+0.8}_{-0.6}\times 10^{9}$\,\Msun\ and
  gas-to-dust ratio of $\approx 50$.  Furthermore, we show that the
  distinct apparent radial profiles of the gas and dust can be
  explained purely by radiative transfer effects (differences in the
  radial optical depths and temperatures) and the observations are
  consistent with the gas and dust mass being similarly distributed
  throughout the starburst.  The latter highlights the importance of
  accounting for radiative transfer effects when comparing molecular
  gas and dust distributions from different tracers.
\end{abstract}

\keywords{Dust continuum emission (412), Galaxy structure (622),
  High-redshift galaxies (734), Millimeter astronomy (1061), James Webb
  Space Telescope (2291)}

\section{Introduction} \label{sec:intro}

Dusty star-forming galaxies represent some of the most intense
starbursts in the universe, making them a unique laboratory to study
the environment for star formation in the early universe
\citep{Casey2014, Hodge2020}.  The spatial distribution of the dust,
gas and stars holds important clues to the triggering mechanism of the
starbursts (e.g., mergers, interactions, accretion events) and, more
generally, provides key constraints for resolved predictions from
theoretical models of galaxy formation
\citep[e.g.,][]{Cochrane2019,Popping2022}.

The advent of sensitive, ground-based sub-mm interferometers (ALMA,
NOEMA) has enabled resolved studies of the dust continuum and
molecular gas emission on (sub)kpc scales in dusty star-forming
galaxies \citep[e.g.,][]{Simpson2015a,Hodge2016,Gullberg2019}.
However, due to their large volumes of gas and dust, the resolved
stellar structure has remained largely obscured.  The latter has now
changed thanks to JWST, which can peer through the dust and spatially
resolve the stellar structure of gas-rich galaxies across the
main-sequence on (sub)kiloparsec scales
\citep[e.g.,][]{Cheng2022,Alvarez-Marquez2023,Colina2023,
  Huang2023,Gillman2023,Gillman2024,Boogaard2024,Hodge2025}.  For
dusty star forming galaxies, resolved studies have shown that the
low-$J$ $^{12}$CO emission tracing the molecular gas is often
extended, similar to the starlight, while the dust continuum is
observed to be more compact \citep{Hodge2015, Chen2017, Calistro2018,
  Tadaki2020, Ikeda2022, Tadaki2023}.

GN20 is a prototypical, massive dusty star forming galaxy at a
redshift of 4.055, about 1.5\,billion years after the big bang.  It is
a bright submillimeter galaxy ($f_{850\,\micron} = 20$\,mJy) with a
total infrared luminosity of about
$L_{\rm IR} = 1.5\times 10^{13}$\,\Lsun\ \citep{Tan2014, Cortzen2020},
located in an overdensity or protocluster \citep{Pope2005, Pope2006,
  Daddi2009, Carilli2011}.  The bulk of the galaxy is completely
obscured at rest-frame UV/optical wavelengths, with only an arc of
UV-bright emission showing at a distance of about 4\,kpc from the
center \citep{Carilli2010, Hodge2012, Hodge2015}.  The molecular gas
is distributed in an extended clumpy structure with a diameter of
$14\pm4$ \,kpc, with a total molecular gas mass of
$M_{\rm mol} = (1.3 \pm 0.4) \times 10^{11} (\alpha_{\rm CO} / 0.8)$
\,\Msun\ \citep[][cf.\ \citealt{Carilli2010}]{Hodge2012} and the
molecular and ionized gas kinematics are consistent with a rotating
disk with $M_{\rm dyn} \approx 5 \times 10^{11}$\,\Msun\
\citep{Hodge2012, Bik2024, Ubler2024}.  High-resolution PdBI
dust-continuum observations at 880\,\micron\ revealed a bright dusty
starburst centered on the molecular gas reservoir and elongated in the
north-south direction \citep{Hodge2015}, but not as extended as the
molecular gas disk, though the extended emission may have been missed
due to limited sensitivity in combination with a dust temperature
gradient.

Recent rest-frame 1.1\,\micron\ imaging with JWST/MIRI
\citep{Colina2023} spatially resolved the stellar structure for the
first time, finding it extends over a similar region as the molecular
gas, with an unresolved nucleus offset by about 1\,kpc from the center
of what appears as a stellar disk with an effective radius of
$r_e=3.6$\,kpc (axis ratio $b/a=0.8$). The offset nucleus suggests
that GN20 is, or has been, involved in an interaction or merger event,
which is not unexpected given it is the brightest galaxy in a
protocluster \citep{Colina2023}.  The total stellar mass of the system
is $M_{*} = (8.6 \pm 4.3) \times 10^{10}$\,\Msun\ with a star
formation rate (SFR) of $2550 \pm 150$\,\Msun~yr$^{-1}$ \citep[][cf.\
\citealt{Tan2014}]{Crespo2024}.  \Paa\ observations with the MIRI MRS
showed a clumpy structure that extends out to about 6\,kpc in radius
\citep{Bik2024}.  The inferred
$\mathrm{SFR}_{\Paa}=205\pm15$\,Msun\,yr$^{-1}$ is only a small
fraction of the total SFR, which implies a large extinction
($A_V = 17.0$ or $A_{V, {\rm mixed}}= 42$; values updated from
\citealt{Bik2024}), even at the wavelength of \Paalpha\
($A_{\Paa}=2.7$).  This is also consistent with the non-detection of
\Pab\ \citep{Bik2024} and the weaker \Ha\ and redder continuum in the
nucleus compared to the UV-bright region from the NIRSpec MSA PRISM
spectrum \citep{Maseda2024}.  NIRSpec IFU observations of \Ha\
\citep{Ubler2024} show emission extended over 15\,kpc, peaking in the
UV-bright region to the north-west.  The large-scale ionised gas
kinematics are consistent with a turbulent, rotating disk, as seen in
the molecular gas \citep{Hodge2012}.  There are some indications that
GN20 may host an AGN: there is weaker, broad \Ha\ emission detected in
the center, coinciding with the peak of the stellar emission, that
could indicate high-velocity outflows or a broad-line region around an
AGN, as well as an elevated \NII/\Halpha\ ratio \citep{Ubler2024}.
PAH emission was detected at 6.2\,\micron\ with Spitzer, consistent
with the large SFR, with the underlying dust continuum consistent with
being powered by a starburst and potentially a faint, dust-obscured
and Compton-thick AGN, that may contribute up to $\sim$15\% of the
infrared luminosity, but remains undetected in deep X-ray observations
\citep{Riechers2014}.  \cite{Cortzen2020} argued that the far-IR dust
emission is best described with a dust SED becoming optically thick at
$\lambda_0 = 170\pm23$\,\micron, yielding a dust temperature of
$52\pm5$\,K, close to the excitation temperature ($48\pm12$\,K) of the
optically thin \CIfirboth\ lines.  GN20 is difficult to observe in
\CIIfsl\ due to poor atmospheric transmission at 376 GHz, but was
recently detected in \NIIfslb\ \citep{Kolupuri2025}.

In this paper, we present new sensitive and high-resolution
$0\farcs13$--$0\farcs23$ (0.9--1.6\,kpc) NOEMA imaging of the 1.1\,mm
(rest-frame 220\,\micron) dust emission of GN20.  We compare this to
JWST/NIRCam, NIRSpec and MIRI imaging and spectroscopy, that now
detects and resolves the stellar and ionized gas emission, as well as
the existing JVLA $^{12}$CO(2--1) and PdBI 880\,\micron\ dust
continuum imaging.  The observations are presented in
\autoref{sec:methods}.  In \autoref{sec:results}, we show that dust is
detected over the full stellar and molecular gas disk of GN20, albeit
with a more concentrated light distribution.  In \autoref{sec:RT
  modeling}, we perform self-consistent radiative transfer modeling of
the integrated CO and dust continuum emission to measure the gas and
dust mass of the system.  We then model the integrated emission
together with the resolved CO and dust emission.  We show that the
distinct observed profiles are in fact consistent with a single gas
and dust distribution throughout the starburst and can be explained
simply due to radiative transfer effects.  The implications of these
results for comparing observed gas and dust profiles and sizes are
discussed in \autoref{sec:discussion}.  The extended dust in GN20
implies a lower dust optical depth than previously reported by
\cite{Cortzen2020}.  Finally, we discuss the mounting evidence that
GN20 experienced a recent interaction or merger that likely
invigorated the starburst.  Throughout this paper, we
adopt %
a \cite{PlanckCollaboration2018a} cosmology (flat $\Lambda$CDM with
$H_{0} = 67.66$\,km\,s$^{-1}$\,Mpc$^{-1}$, $\Omega_{m} = 0.3111$ and
$\Omega_{\Lambda} = 0.6889$).  %
At the redshift of GN20 ($z=4.055$) $1\farcs0$ corresponds to a proper
distance of 7.07\,kpc. We use $\log$ to denote $\log_{10}$ and $\ln$
for the natural logarithm.

\section{Observations and data reduction} \label{sec:methods}
\subsection{NOEMA 1.1\,mm}
\label{sec:noema-data}
\begin{figure}[t]
  \centering
  \includegraphics[width=\columnwidth]{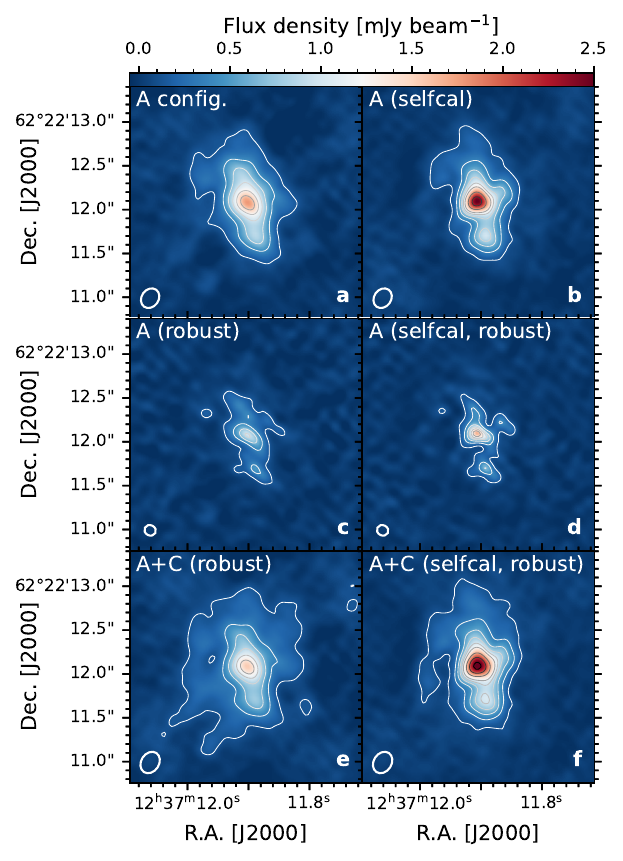}
  \caption{\label{fig:cutout_weighting} IRAM/NOEMA long-baseline
    1.1\,mm imaging of GN20.  The left and right columns show the
    cleaned images before and after self calibration.  The panels show
    the (a) highest resolution A-configuration tracks only, at natural
    resolution ($0\farcs24\times0\farcs2$), before and (b) after self
    calibration, (c) A tracks, robust weighted to higher resolution
    ($0\farcs14\times0\farcs12$), before and (d) after self
    calibration, (e) A+C configuration tracks, robust weighted to the
    A configuration-only resolution ($0\farcs26\times0\farcs2$) before
    and (f) after self-calibration.  Contours are overlaid on all
    images, starting at $5\sigma$ increasing in steps of $10\sigma$ up
    to $55\sigma$ and then steps of $20\sigma$.}
\end{figure}
GN20 was observed with NOEMA in A configuration using all 12 antennas
with the longest baselines up to 1.7\,km.  The band 3 receivers were
tuned to 261\,GHz (1.1\,mm) using the Polyfix correlator with 16\,GHz
of total bandwidth in the upper and lower sideband.  Four tracks of
varying length were obtained during the winter semester of 2023, on
February 21st and 23rd and March 1st and 4--5th.  The first three
tracks suffered from poor observing conditions and high phase rms, but
the final track was taken under good conditions and provided a total
of 2.3\,h of on-source data.  Given the limited $uv$-coverage of this
single track and the thus asymmetric beam, two additional tracks were
observed in A configuration during the winter of 2024 on February 17th
and March 7th.  The former track again suffered from poor observing
conditions, while the final track was taken under better weather and
provided 2.6\,h of on-source time with orthogonal $uv$-coverage.  In
addition to the A configuration data, a 2.2\,h track in C
configuration was observed on April 19--20th, 2023 to fill in short
spacings.

The data reduction and imaging were performed at IRAM using
\textsc{gildas}\footnote{\url{http://www.iram.fr/IRAMFR/GILDAS}}.  The
bandpass calibrators were 3C273 (March 4 and April 19) and 3C279
(March 7) and the flux calibrators were J2010+723 (March 4) and MWC349
(April 19 and March 7). J1125+596 and J1302+690 were used as amplitude
and phase calibrators on all tracks (using average polarisation mode,
as both sources are polarised). %
We inspect the cubes for parasites and ranges with elevated rms and
mask channels accordingly.  We then combine the sidebands adopting the
measured continuum spectral index between the sidebands of -3.5.

We imaged both the A tracks individually, providing the highest
resolution data, as well as the combination of the A and C tracks.
Only the two good A configuration tracks were used in the end, as we
found that adding in the additional, low quality, A configuration
tracks did not impact the final images given the low weight of the
visibilities.

We clean the maps down to their $1\sigma$ noise level using an
elliptical $2\farcs2\times1\farcs7$ mask centered on the source at a
matching position angle using the \textsc{HOGBOM} algorithm (using
multi-scale clean instead yields essentially the same map).  The
synthesized beam of the naturally weighted image using just the A
tracks is $0\farcs25\times0\farcs2$.  We also apply robust weighting
to obtain a map at very high ($0\farcs14\times0\farcs12$) resolution,
highlighting the small-scale structure.  Both images are shown in
\autoref{fig:cutout_weighting}.

As the cleaned images still show structure in the residuals that could
indicate poor phase solutions, we subsequently self-calibrate the
data.  We use three iterations of phase self calibration with
integration times of 135, 90 and 45 seconds (the time per scan is 22
seconds), increasing the number of clean components to build the
source model as 50, 100, 150, using the same clean mask.  The number
of clean components is motivated by the number required to describe
the source structure without introducing negative clean components.
We let the minimum signal-to-noise ratio for a visibility to be
self-calibrated be 3 and keep the baselines that cannot be
self-calibrated.  We experimented with varying the averaging times,
number of clean components and the size of the clean support within
reasonable limits, and found that this has limited impact on the final
result, yielding qualitatively the same map.

We perform the same procedure for the combined A+C configuration data.
The root-mean-square noise in the map is 23\,$\mu$Jy\,beam$^{-1}$ (for
comparison, the theoretically expected rms is
21.8\,$\mu$Jy\,beam$^{-1}$).  As the naturally weighted beam for the
combined data is larger ($0\farcs35\times0\farcs3$), we apply robust
weighting to achieve approximately the same resolution as the A
configuration data for the final map.  The self-calibrated maps are
shown in \autoref{fig:cutout_weighting}.  The beam size of the final
map is $0\farcs26\times0\farcs2$ corresponding to a physical
resolution of $1.77\times1.41$\,kpc, with an rms noise of
21.5\,$\mu$Jy\,beam$^{-1}$.

Comparing total fluxes in the maps in \autoref{fig:cutout_weighting}
row by row, we find roughly 10.0, 9.0 and 11.5\,mJy, with similar
results before and after selfcal.  This implies most of the flux is
already recovered in the A-configuration map alone, with the
high-resolution map missing a larger fraction of the extended flux,
while the final map including the low-resolution C-config data
recovers all the flux (see \autoref{sec:results}).

\begin{figure*}[t]
  \centering
  \includegraphics[width=\textwidth]{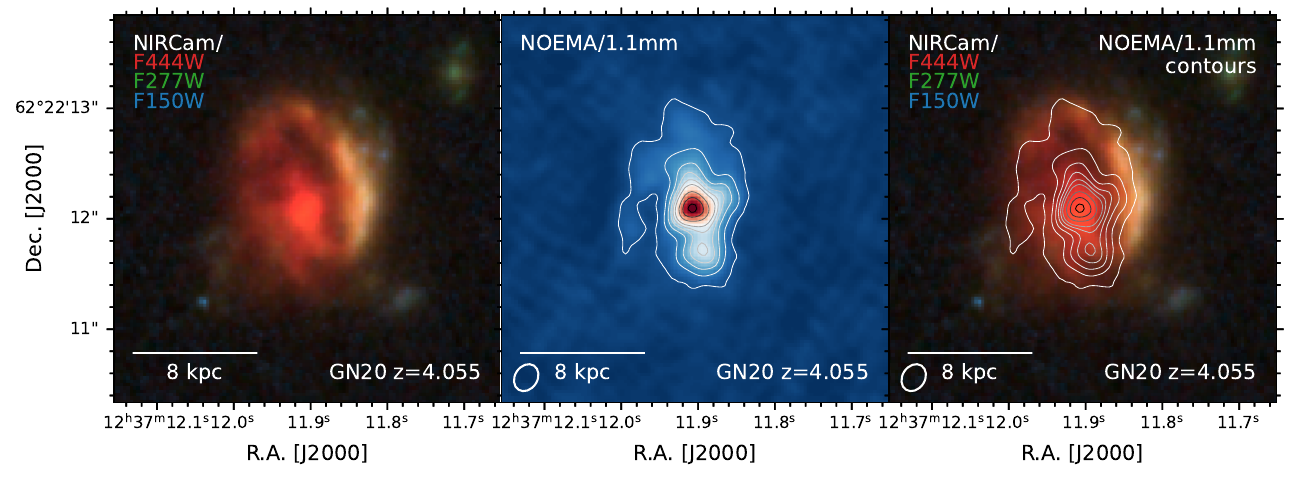}
  \caption{\label{fig:rgb_noema} JWST/NIRCam and NOEMA 1.1\,mm imaging
    ($4'\times4'$ cutouts) of GN20 at $z=4.055$.  Dust contours start
    at $5\sigma$ and increase in steps of $10\sigma$ up to $55\sigma$
    and then steps of $20\sigma$ up to $115\sigma$.}
\end{figure*}
\subsection{Ancillary resolved data}
\label{sec:ancillary-data}
The dust continuum at 880\,\micron\ (rest-frame 175\,\micron) was
obtained with the Plateau de Bure Interferometer and presented in
\cite{Hodge2015}.  These data have a synthesized beam of
$0\farcs3\times0\farcs2$, PA=37$^{\circ}$ and an rms of
0.25\,mJy\,beam$^{-1}$ (about 10$\times$ higher than the NOEMA data
presented here).

The $^{12}$CO(2--1) imaging (hereafter CO) was obtained with the
Jansky Very Large Array at 45\,GHz and we use the moment 0 map
presented by \cite{Hodge2012}, with a synthesized beam size
$0\farcs19\times0\farcs17$.

JWST/MIRI and NIRCam imaging of GN20 were taken as part of program ID
1264 (PI: Luis Colina).  The MIRI/F560W imaging of GN20, tracing the
rest-frame $\sim$1\,\micron\ stellar emission is presented in
\cite{Colina2023}.  The data processing of the complete MIRI imaging
of GN20 and known members of the protocluster (in F560W, F770W, F1280W
and F1800W) is presented in \cite{Crespo2024}.  For NIRCam, F200W and
F356W imaging with 1825.3s of integration time were distributed in
five dithers, each with one integration, using seven groups with the
SHALLOW4 readout mode.  Imaging in F115W, F150W, F277W and F444W used
the same readout configuration for a total time of 2555.4 seconds
distributed in seven dithers.  The NIRCam data was calibrated using a
custom procedure built on the JWST calibration pipeline (v1.12.3) with
CRDS context 1145. The process includes the removal of snowballs and
wisps following \cite{Bagley2023}, and a superbackground
homogenization with 1/f correction as described by
\cite{Perez-Gonzalez2023}. The final mosaics were resampled to a
uniform pixel scale of 0.03 arcsec/pixel.

The JWST/MIRI MRS map of \Paalpha\ is presented in \cite{Bik2024}.
The \Halpha\ map from the JWST/NIRSpec IFU is presented in
\cite{Ubler2024}.

\begin{figure*}[t]
  \centering
  \includegraphics[width=1.0\textwidth]{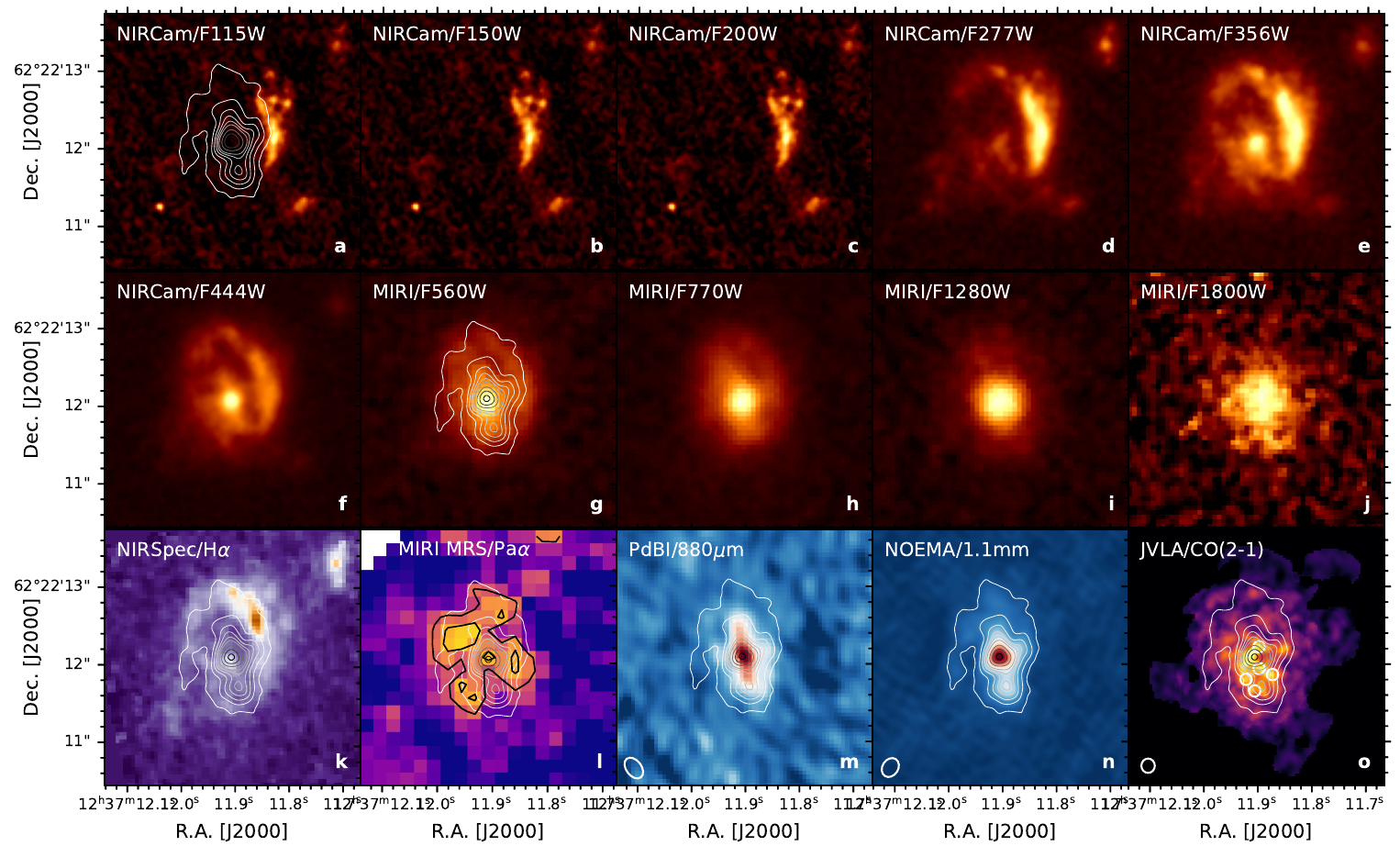}
  \caption{Multi-wavelength cutouts of GN20.  The top two rows show
    the NIRCam and MIRI imaging (\citealt{Colina2023, Crespo2024})
    covering the rest-frame 0.23\,\micron--3.56\,\micron\ range.  The
    bottom row shows the \Ha\ map \citep{Ubler2024}, \Paa\ map
    \citep[][with 2--3$\sigma$ contours overlaid]{Bik2024}, the dust
    continuum at 880\,\micron\ \citep[PdBI][]{Hodge2012} and 1.1\,mm
    (NOEMA; this work) and the JVLA/CO(2--1) \citep[][with their CO
    clumps indicated]{Hodge2012}.  All maps are shown on a linear
    colorscale and NOEMA contours are overlaid on selected images,
    starting at $5\sigma$ and increasing in steps of $10\sigma$ up to
    $55\sigma$ and then steps of $20\sigma$ up to $115\sigma$.  Beam
    sizes of the radio maps are shown in the bottom left
    corner. \label{fig:cutouts}}
\end{figure*}

A JWST/NIRCam color image of GN20 is shown in comparison to the new
NOEMA data in \autoref{fig:rgb_noema}.  Multi-wavelength cutouts are
shown in \autoref{fig:cutouts}.

\section{Analysis and Results} \label{sec:results}

\subsection{Dust emission at rest-frame 220\,\micron}
\label{sec:dust-emission}
The dust emission from GN20 at 1.1\,mm (rest-frame 220\,micron) is
well resolved and extended, and detected out to the same radius as the
rest-frame 1--2\,\micron\ stellar emission (traced by MIRI) and the
molecular gas emission (traced by CO).  The total dust emitting region
spans an elliptical region of around $13.8 \times 10.3$\,kpc
($1\farcs95\times1\farcs45$) at $3\sigma$, or $12.4 \times 8.8$\,kpc
($1\farcs75 \times 1\farcs25$) at 5$\sigma$.

The total flux in the final map is
$f_{\nu, 1.1\,{\rm mm}} = 11.45 \pm 0.23$\,mJy, %
which is consistent with the total flux
from the C-configuration track and the flux measured at 1.1\,mm by
AzTEC at the JCMT of $11.45\pm0.99$\,mJy \citep{Perera2008}.

The 1.1\,mm dust emission extends asymmetrically around the nucleus in
the north-south direction.  A significant clump is detected towards
the south, and extended emission is also seen to the east.  The
highest resolution map reveals several fainter peaks in the dust
emission at these locations (see \autoref{fig:cutout_weighting}),
implying a clumpy structure in the dust on these scales.  On larger
scales, there is fainter emission extending in all directions, but in
particular towards the north, over the full extent of the stellar and
molecular gas in GN20 (\autoref{fig:cutouts}).

Of the total flux, $\approx$2.6\,mJy is contained in the central point
source at (23\%) and $\approx$1.0\,mJy in the southern point source
(9\%).  The total flux contained in the brightest central region of
$\approx 0\farcs5\times0\farcs2$ along the major axis of the source is
$\approx$7\,mJy (measured as the flux with in the $20\sigma$\,contour)
indicating that a significant fraction of the total emission is coming
from large radii.

While the continuum morphology of GN20 is complex, to quantify the
overall extent of the 1.1\,mm dust emission and facilitate a
comparison to other work, we fit a two-dimensional Gaussian as well as
a \cite{Sersic1968} profile to the cleaned image using \textsc{galfit}
\citep{Peng2002}.  We use the Gaussian synthesized beam model as
kernel to obtain `deconvolved' size measurements.  The best-fit
Gaussian model yields a FWHM of the major axis of
$0\farcs64 \pm 0\farcs01$ ($4.53\pm0.02$\,kpc) with a
$b/a = 0.488 \pm 0.003$.  We fit both a model with free S\'ersic index
$n$, as well as a model with $n=1$.  The former provides a best-fit
for $n=2.01\pm0.03$, with an effective radius of $3.18\pm0.04$\,kpc
($0\farcs45$) and axis ratio $b/a=0.458\pm0.004$.  Similar results are
obtained for the $n=1$ fit, with $r_e = 2.48 \pm 0.02$\,kpc
($0\farcs35$) and $b/a = 0.501 \pm 0.003$.  Note that the
uncertainties are typically underestimated by \textsc{galfit} (see
\citealt{vanderWel2012}).  Overall we conclude that
$r_{e}({\rm 1.1\,mm}) \approx 2.5$\,kpc ($0\farcs35$--$0\farcs4$) with
an axis ratio $b/a\approx0.5$.

\subsection{Comparison to multi-wavelength data}
\label{sec:comp-multi-wavel}
A multi-wavelength overview of GN20 is shown in \autoref{fig:cutouts}.
The bulk of the dust emission is coming from the region that is
completely obscured in bands bluer than F277W (rest-frame
0.55\,\micron), and extends up to the edge of the UV-bright region in
the west.  This is consistent with dust obscuring most of the galaxy
in the rest-frame UV/optical.  The dust emission peaks at
R.A.=12$^{\rm h}$37$^{\rm m}$11$^{\rm s}$906,
Decl.=+62$^{\circ}$22'12\farcs091.  The peak is co-spatial with the
peak of the stellar emission at rest-frame 1.1\,\micron\ (traced by
MIRI) and the molecular gas emission (traced by CO), and has a similar
extent.

The NIRCam imaging now clearly reveals the morphology of the arm-like
feature that loops around the western part of the galaxy from the
south to north, of which the least-obscured part coincides with the
UV/optical-bright part of the galaxy.  The `gap' in the stellar
emission between the nucleus and arm to the north-west already seen in
MIRI aligns with a drop in the dust emission, suggesting this is a
real gap in the stellar distribution separating a spiral arm or tidal
feature, rather than being caused by spatially varying extinction.
The clump in the dust emission to the south is coincident with one of
the clumps identified in the 1.1\,\micron\ residual image by
\cite{Colina2023} (cf.\ \citealt{Crespo2024}).

The north-south bar feature in the dust emission matches that seen at
880\,\micron\ \citep{Hodge2015}.  A similar structure is now also seen
in the rest-frame near-infrared stellar morphology around
$\sim$1--1.5\,\micron\ and indicates the presence of a stellar bar in
GN20.  The 1.1\,mm-clump in the south then roughly coincides with
enhanced star formation at the tip of the bar.  A detailed study of
the stellar bar will be presented in a forthcoming paper.  Note most
of the smaller-scale clumps and structural details seen at 1.1\,mm are
not visible in the early (6 antenna) PdBI map at 880\,\micron, likely
because of the $10\times$ higher rms noise.

The central region that is brightest in dust continuum emission is
largely co-spatial with the region brightest in CO and the peak of the
dust emission is close to the peak of the CO emission.  The dust clump
to the south is not co-spatial with, but rather in-between, the CO
clumps identified by \citet[][indicated on
\autoref{fig:cutouts}]{Hodge2012}.  Interestingly, the \Paa\ emission
\citep{Bik2024} seems to avoid the regions that have the highest dust
intensity, in particular the bright clump in dust emission towards the
south, instead being brightest in the north-east where the dust
emission is fainter and more extended.  The \Paa\ peak in the center
is cospatial with the dust emission.  A similar picture is seen in
\Ha\ \citep{Ubler2024}.  While there does not need to be one-to-one
correspondence between the extincting and emitting dust, this picture
is consistent with higher extinction in those regions.

\subsection{Radial profiles}
\label{sec:radial-profiles}
\begin{figure}[t]
  \centering
  \includegraphics[width=\columnwidth]{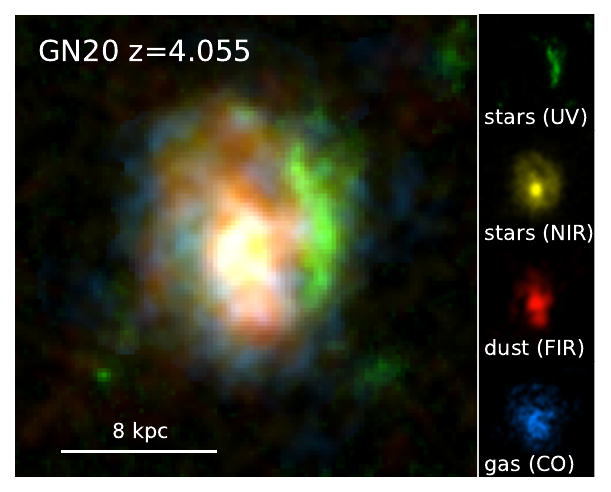}
  \caption{\label{fig:false_color} $3''\times3''$ false color image of
    GN20 at $z=4.055$ highlighting the distinct spatial distribution
    of the different components: rest-frame UV (NIRCam/F150W),
    rest-frame near-IR (MIRI F560W, yellow), rest-frame 220\,\micron\
    dust continuum (NOEMA/1.1mm) and the molecular gas traced by
    CO(2--1) (VLA).  The stellar distribution traced by the rest-frame
    near-IR is largely cospatial with the CO and the dust emission,
    though the emission from the latter appears more centrally
    concentrated.  The (unattenuated) rest-frame UV light is only
    visible in the outskirts.}
\end{figure}

\begin{figure}[t]
  \centering
  \includegraphics[width=1.0\columnwidth]{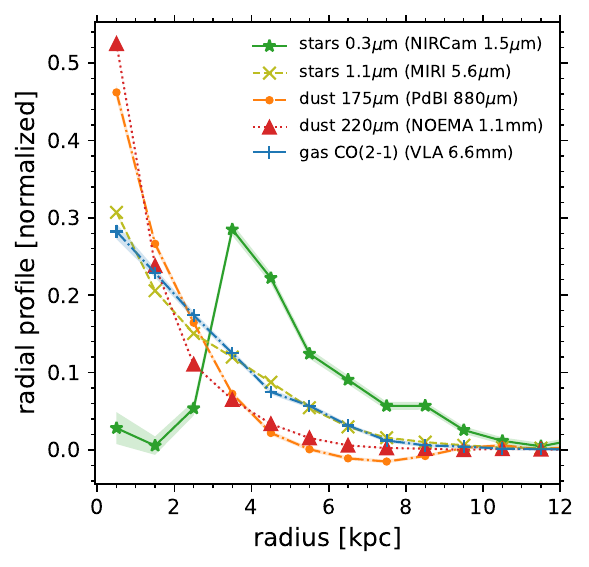}
  \caption{\label{fig:radial-profiles} Azimuthally averaged and
    integral-normalized radial profiles, measured from the peak of the
    NOEMA continuum emission in bins of 1\,kpc.}
\end{figure}

A false color-image illustrating the spatial distribution of the
starlight in the rest-frame UV and near-IR, as well as the dust and
gas is shown in \autoref{fig:false_color}.  To study the spatial
distribution of the stellar, dust and gas emission, we compute
azimuthally averaged radial profiles using \textsc{photutils},
propagating errors.  We use circular profiles, to facilitate a
comparison given the different axis-ratios between the dust, stars and
gas in GN20; with a lower ellipticity in the stellar emission, and
similar or even lower for the gas.  The resulting profiles are shown
in \autoref{fig:radial-profiles}.

Notably, the rest-frame near-IR stellar emission and the molecular gas
follow each other closely, at all radii.  In contrast, the dust
emission both at 1.1\,mm (rest-frame 220\,\micron) and (at lower S/N)
880\,\micron\ (rest-frame 175\,\micron), shows a significantly more
concentrated distribution.  This is consistent with the
two-dimensional S\'{e}rsic fits, that yield a smaller effective radius
for the dust ($r_{e, 220\,\micron} \approx 2.5$\,kpc) compared to the
stellar disk ($r_{e, 1.1\micron} = 3.6$\,kpc; \citealt{Colina2023}).
The rest-frame UV emission instead is offset with a peak around 4\,kpc
and little emission from the center.  Note that due to the asymmetric
dust emission and circular apertures, more dust emission appears to
come from the UV-bright region than is actually the case
(cf. \autoref{fig:cutouts}).

\section{Radiative Transfer modeling}
\label{sec:RT modeling}

\begin{figure*}[t]
  \centering
  \includegraphics[width=1.0\textwidth]{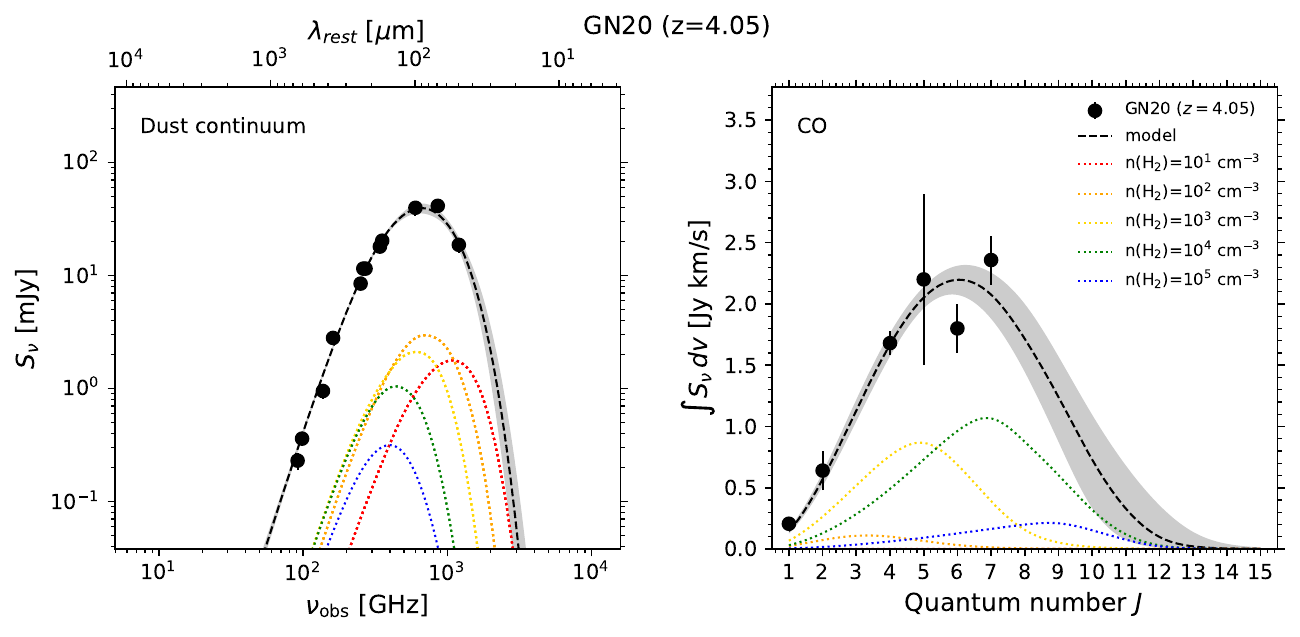}
  \caption{\label{fig:turb_fit} Best-fit TUNER model for the
    integrated (unresolved) GN20 data.  The black points show the
    observations, the dashed line with gray band the median model with
    $\pm 1\sigma$ percentiles.  The colored lines highlight the
    contributions of a few individual density components (scaled up
    for visibility).  Overall the model provides a very good
    simultaneous fit to the dust continuum and CO data.}
\end{figure*}

A key question that arises from \autoref{fig:false_color} and
\autoref{fig:radial-profiles} is whether the distinct radial profiles
of the gas and the dust imply that the gas and dust have intrinsically
different spatial distributions, or whether they could also arise from
gas and dust with the same spatial distribution, with the observed
different profiles being driven by radiative transfer effects
\citep[e.g.,][]{Calistro2018}.

To investigate this, and also to derive the molecular gas and dust
masses and properties of GN20, we model the CO and the dust continuum
emission using a radiative transfer model that we call TUNER
(Turbulent Non-Equilibrium Radiative Transfer).  TUNER uses a
physically-motivated (lognormal) density and density-coupled
temperature distribution, together with the Large Velocity Gradient
(LVG) and escape probability approximations, to solve for both the
line and continuum opacities of the molecular ISM simultaneously and
self-consistently.  The details of the TUNER model are explained in
more detail in \autoref{sec:tuner}.

We first use TUNER to model the integrated (unresolved) CO and dust
continuum SED of GN20 in \autoref{sec:turb}.  We then couple
individual TUNER models to a S\'{e}rsic profile and a radial
temperature distribution to study the resolved CO and dust profiles of
GN20 in \autoref{sec:ted}.  The new and existing far-infrared
photometry and line-flux measurements for GN20 are compiled in
\autoref{tab:fluxes}.

\subsection{Integrated CO and dust SED} \label{sec:turb}
The integrated CO and dust continuum emission from GN20 is shown in
\autoref{fig:turb_fit}.  For the CO(2--1) flux of GN20, we adopt the
lowest-resolution integrated flux measurement from \cite{Carilli2010},
which is lower than that from \cite{Carilli2011} and \cite{Hodge2012},
though consistent within error, but best-agrees with the results from
the radiative transfer modeling, as discussed below.

For the TUNER modelling, we adopt uniform priors on the parameters
with the following bounds: radius $\in [0.1, 10^{4}$]\,pc,
$n_{\rm H_2}$ $\in [1, 10^7]$\,cm$^{-3}$, $\Tkin \in [10, 600]$\,K, a
temperature power law slope as function of density
$\gamma_{T} \in [-0.5, 0.05]$, abundance
$\COHH \in [10^{-6}, 10^{-3}]$, gas-to-dust ratio $\gdr \in [1, 250]$,
dust opacity slope $\beta \in [1.2, 3.6]$.  The prior on the dust
temperature is coupled to the \Tkin\ such that
$\Tkin/\Tdust \in [0.5, 6.0]$. We assume a virial parameter
$\kappa_{\rm vir} = 1$ and a dust opacity of
$\kappa_{0} = 0.47$\,cm$^{2}$\,g$^{-1}$ at $\nu_{0} = 352.7$\,GHz
\citep{Draine2014}.

A key parameter is the turbulent velocity width of the lognormal
distribution, \dvturb, which turns out to be poorly constrained in the
unresolved modeling of GN20 (likely because of the lack of
high-density tracers), and asymptotes to high \dvturb\ values
($\gg$100\,\kms), with a weak local minimum around $\sim$20--40\,\kms.
While these broad distributions provide equally good fits, but may
have unrealistically large masses, because they can hide a lot of
material at high density and low temperature where it emits very
little.  We therefore constrain the turbulent width in the model using
the resolved spectroscopic data.  \cite{Hodge2012} measured the width
of individual CO clumps in the disk to be 90--150\,\kms, with the
central and highest-S/N clump having a $\mathrm{FWHM} = 90\pm40$\,\kms
(after rotation and beam-smearing corrections).  Given the beam size,
these line widths are still upper limits on the actual velocity
dispersion in the gas.  For the ionized gas, which typically has a
dispersion of a factor 2--$3\times$ larger than the cold gas
\citep{Girard2021, Rizzo2024}, \cite{Bik2024} find a dispersion of
40--140\,\kms\ in \Paa, while \cite{Ubler2024} find $90\pm10$\,\kms\
for \Ha.  From the resolved modeling (see \autoref{sec:ted}) we
consistently find \dvturb\ in the range of 20--65\,\kms, with a
best-fit value around 30\,\kms.  We therefore fix
$\dvturb = 30$\,\kms, but also perform two runs with 90 and 10\,\kms\
(smaller values \dvturb\ require unrealistically large radii of
$>10$\,kpc to match the observed fluxes), finding consistent results.

The best fit is shown in \autoref{fig:turb_fit} and the full posterior
in \autoref{fig:turb_corner} in \autoref{sec:corners}.  Overall, the
model provides a very good description of the unresolved CO and dust
SED of GN20.  To show how the emission of different density components
in the model contributes to the overall SEDs, we highlight a subset of
the components (by powers of 10, scaled up for clarity).  The dominant
contributions to the CO ladder come from \HH\ gas at densities around
$10^{3}$ and $10^{4}$ cm$^{-3}$.  Note there is observational tension
between the CO(6--5) and CO(7--6) measurements which cannot both be
fit by the model simultaneously.  This is most likely a limitation of
the data, as it is difficult to imagine a physical scenario in which
the CO ladder has a sudden `dip' at the CO(6--5) transition, unless
there is an additional component that strongly picks up in the
high--$J$ lines.  Note if we remove either transition from the data we
can provide an excellent fit to the remaining data points, with little
impact on the total mass or conclusions of the paper.  Future
observations of higher-$J$ lines could shed light on this.

From the unresolved model, we derive a total molecular gas mass of
$\Mmol = 3.4_{-1.8}^{+5.9} \times 10^{11}$\,\Msun, with an
$\aco = 3.6_{-1.4}^{+7.3}$, higher than the dynamically constrained
limit from $1.1\pm0.6$ from \citep[][though consistent within
substantial errors]{Hodge2012}.  The larger upward errors are partly
driven by the high turbulent velocity width, that implies a
significant fraction of high-density gas that is hard to constrain
with the unresolved $^{12}$CO data alone, though we refer to
\autoref{sec:ted} for more precise constraints when including resolved
data (in good agreement with the median values).

At face value the molecular gas mass estimate is higher than the gas
mass estimate from the single CO(1--0) line of
$1.3 \pm 0.4 \times 10^{11} (\aco/0.8)$ \Msun\ \citep{Carilli2010},
though it is slightly lower after accounting for the difference in
conversion factors, as the model predicts a slightly lower CO(1--0)
flux.  Earlier single- and two-component LVG modelling by
\cite{Carilli2010} noted that the CO(1--0) flux could not be described
by a single gas component (underpredicting the flux by a factor
$\geq 2\times$) and required a second, more extended, low density
component (it should be noted that these two component models are
highly degenerate and strong assumptions are required on the sizes of
these components; cf. \citealt{Daddi2015}).  We now find much better
agreement using the turbulent gas distribution (with fewer free
parameters), but note that the measured CO(1--0) flux from
\cite{Carilli2010} is still underpredicted by $\sim$30--50\% (at
$1.5\sigma$ significance).  While this could simply be the result of
the measurement error, in particular on the faint low-$J$ CO lines,
given the good overall agreement for the higher-$J$ lines, together
with the potentially higher CO(2--1) flux, this could also imply the
presence of an additional gas component with a significantly distinct
spatial distribution (such as a low-density component with a much
larger extent radius than that occupied by the higher-density gas),
such that it cannot be captured through a single lognormal
distribution.  This kind of distinct two-component (turbulence)
modeling is significantly more degenerate and thus requires additional
assumptions which we leave to future work.  Instead, we focus on
modeling the resolved profiles together with the integrated values in
the next section \autoref{sec:ted}.  Still, to test the impact on the
gas-mass estimate of a potential low-density, low-excitation
component, however, we run an additional model using a significantly
higher CO(2--1) flux of $1.0\pm0.3$\,Jy\,\kms
\citep[cf.][]{Hodge2012}, to see if this results in a model with
higher low-$J$ line luminosities.  It does not.  Instead, it hardly
produces a better fit to the CO(1--0) data and a poor fit the
$J_{\rm up}=2$ line, while the total gas mass estimates remain
consistent with the fiducial model.

The estimated total molecular gas mass implies that GN20 is
predominantly gas rich, with the molecular gas mass exceeding the
estimated stellar mass ($\approx 10^{11}$\,\Msun) and comprising the
majority of the total (dynamical) mass of
$M_{\rm dyn} \approx 5 \times 10^{11}$\,\Msun \citep{Hodge2012,
  Ubler2024}.

Turning to the dust, we measure an
$\LIR = 1.6^{+0.1}_{-0.1}\times 10^{13}$\,\Lsun, in perfect agreement
with general opacity model from \citep{Cortzen2020}.  An interesting
result from the self-consistent modeling is that the gas-to-dust
ratio, \gdr, is quite low, with a median posterior value of
$\approx 50$ (about 3x lower than the typical ISM value), while the
median $\COHH \approx 10^{-4}$ is closer to the typical abundance of
CO in the ISM \citep{Frerking1982, Sofia2004, Lacy2017}.  We note that
\COHH\ runs into the upper bound of the prior (set at
$\log(\COHH) = -3.0$), which may impact the median posterior value.
We argue this is not an issue, given that the upper bound is
physically motivated, being about $5\times$ higher than the highest
abundance measured in the Milky Way \citep{Sofia2004, Lacy2017}.
Indeed, taking the carbon abundance from \cite{Asplund2009}, assuming
most hydrogen is in \HH\ at the molecular cloud densities of the
model, and furthermore assuming that all carbon is locked in CO, would
result in $\log(\COHH) = -3.3$.  The latter assumption is certainly a
strong overestimation given that a significant fraction of the carbon
will be locked in dust grains (as said, the measured value in the
solar neighborhood is closer to $-4.0$) and thus especially for GN20,
given its significant dust mass.  Under these assumptions, the adopted
bound of $-3.0$ corresponds to taking about twice the
\cite{Asplund2009} abundance (i.e., further assuming that GN20 has
$\sim$2$\times$ solar metallicity).

The relatively small G/D ratio implies a relatively large fraction of
mass in dust grains, as may be expected in denser environments
\citep[e.g.,][]{Patra2025}.  Alternatively, if the true \gdr\ is
actually solar (i.e., 100-150), given that the \gdr\ is inversely
proportional to the dust opacity (at fixed dust mass), it would imply
that the dust opacity in GN20 is $\sim$2--$3\times$ larger than the
adopted value from \cite{Draine2014}.  Such opacity would be higher
even than the dust opacity from \cite[][used by e.g.,
\citealt{DaCunha2015}]{Dunne2000}, which is a factor $1.8\times$
higher than the \cite{Draine2014} value.  The median posterior
$\beta \sim 2.2$ is higher than the commonly assumed value of 1.8, yet
closer to the value of 2.08 from \cite{Draine2014}.  The total derived
dust mass is $\Mdust = 5.9_{-1.0}^{+0.6} \times 10^{9}$\,\Msun, higher
than the estimate of $2.0\pm0.8 \times 10^{9}$ for optically thick
dust from \cite{Cortzen2020}, consistent with the lower \gdr.  Note
that rescaling the dust opacity in the model does not impact the
derived (dust) mass, as the \gdr\ compensates accordingly (but
consistent dust opacities and $\beta$'s should be used when comparing
mass estimates from different sources).  It should be noted the CO
abundance and \gdr\ do not directly couple and are non-linearly
dependent on the gas conditions and metallicities
\citep{Sandstrom2013, Remy-Ruyer2014}.  As such, the discrepancy
between the \gdr\ and the \COHH\ does not necessarily imply that the
dust opacity is different.  A different dust opacity could point to
significantly different properties of the dust in GN20.  Given the
extreme gas densities and conditions in the dusty starburst galaxy,
this may not be surprising.

\subsection{Resolved CO and dust profiles} \label{sec:ted}

\begin{figure*}[t]
  \centering
  \includegraphics[width=\textwidth]{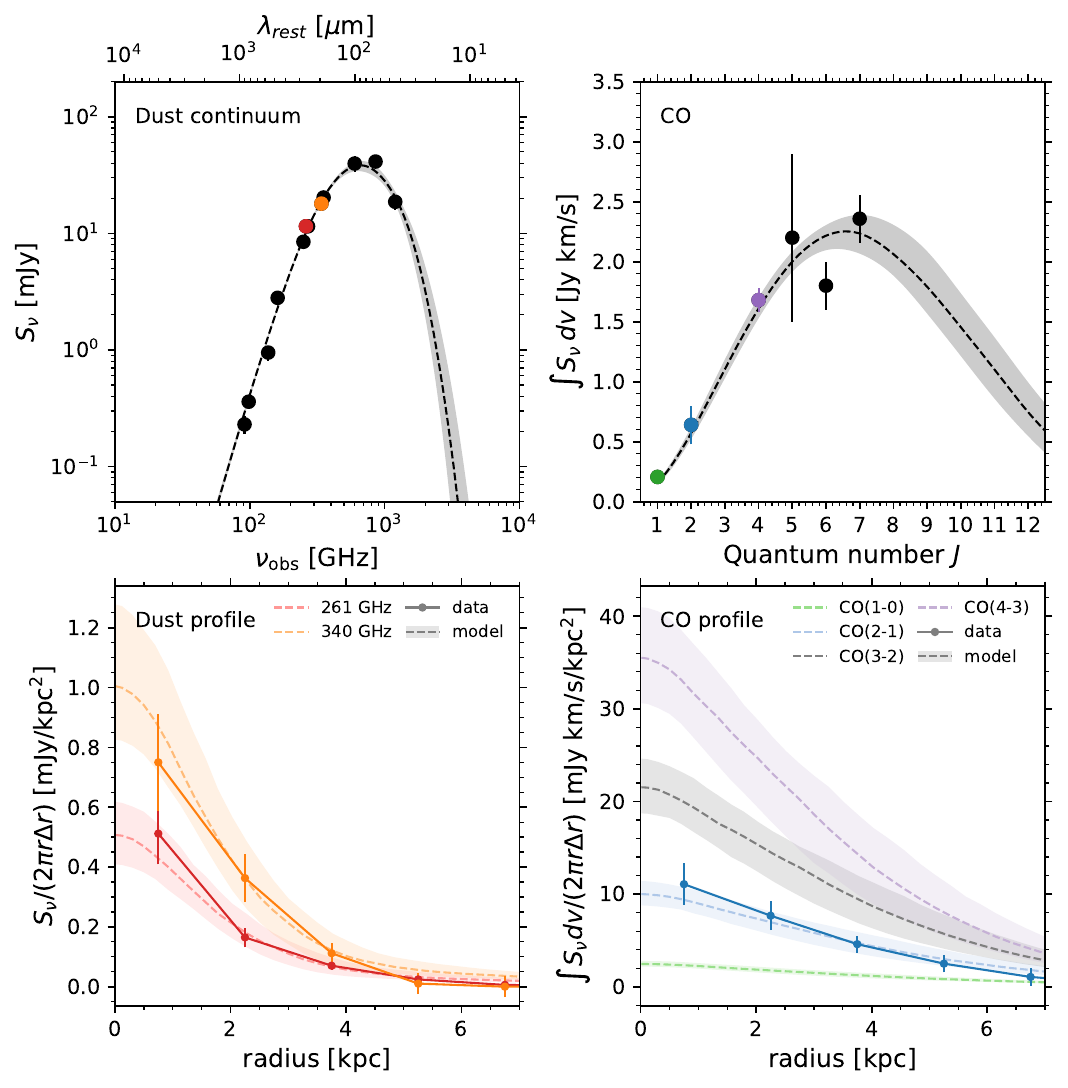}
  \caption{\label{fig:ted_fit} Best-fit TED model for the combined
    resolved and integrated (unresolved) GN20 data. The top panels
    show the integrated dust continuum and CO spectral energy
    distributions, while the bottom panels show the spatially resolved
    dust and CO radial profiles.  The points indicate the data while
    the dotted line with shading the posterior model.  Matching colors
    are used in the top and bottom panels for dust and CO observations
    with both integrated and resolved measurements (and their models).
    The TED radiative transfer model, that is based on a single
    S\'{e}rsic gas mass surface density profile with linear radial
    temperature gradient and constant CO abundance and gas-to-dust
    ratio, can provide a good fit to both the resolved and integrated
    data, implying that the different observed profiles are consistent
    with a single underlying gas- and dust mass distribution.}
\end{figure*}

We model the resolved radial profiles of the CO and dust continuum by
combining individual TUNER models to build a spatially resolved
profile.  We call this combination of TUNER models the TED model
(Turbulent Extended Distribution).  \emph{Critically}, we take a
constant [CO/\HH] and \gdr\ throughout the entire profile, such that
the gas and dust follow the same intrinsic radial distribution.  We
assume the gas column density follows a \cite{Sersic1963} profile with
a linear radial kinetic- and dust temperature gradient ($dT/dr$).  We
divide the profile into annuli and at each radius compute a TUNER
model for the corresponding solid angle of the annulus and mean
density and temperatures.  The density and temperature parameters are
taken to be the values in the center.  The resulting model fluxes of
each ring are both summed to yield total fluxes, to be fit to the
unresolved data, as well as combined into a radial profile and
convolved by the beam of the respective observations, to be fit the
observed profiles.  The model is computed outward in radial bins in
steps that ensure sufficient resolution in density and temperature
until the flux in the CO ground state is converged to within 1\%.  We
model the observed profiles out to a radius of 8\,kpc.  Specific care
needs to be taken to ensure consistency between the unresolved and
resolved flux measurements.  While the total fluxes from the resolved
and integrated observations are generally consistent within error
(cf. \citealt{Hodge2015}), absolute differences will lead to
inconsistencies in the modelling, where the profiles by construction
integrate to the total flux.  To ensure consistency, we adopt a 20\%
calibration uncertainty, an error floor on the error beyond the
effective radius, and renormalize the resolved CO(2--1) and PdBI
850\,\micron\ profiles such that they integrate to their unresolved
integrated flux measurements (mainly relevant for the CO map).

We adopt uniform priors on the scale radius
$r_{\rm exp} \in [500, 6000$]\,pc, S\'{e}rsic index $n \in [0.5, 4]$,
$dT/dr \in [-50, 50]$\,K\,kpc$^{-1}$ and model all other parameters
with the same priors as in the unresolved modeling
(\autoref{sec:turb}), with two changes: we leave \dvturb\ free, with a
prior of $\dvturb \in [1,90]$ (see \autoref{sec:turb}), as we find
that it can be well constrained from the TED model.  In contrast, we
find that leaving the abundances free provides very degenerate
constraints with solutions that have several unphysical
characteristics (including large gas(-to-dust) masses and conversion
factors, that exceed the dynamical constraints by $\geq$10$\times$).
We therefore fix the abundances to values close to the median best-fit
values from the unresolved TUNER model ($\gdr=50$ and
$\COHH = 10^{-4}$).  Following the discussion in \autoref{sec:turb},
we also run a model with $\COHH = 10^{-4.3}$ (closer to the posterior
peak) and corresponding $\gdr=65$ (note their non-linear degeneracy in
\autoref{fig:turb_corner}), finding the same results.

The results from the TED model are shown in \autoref{fig:ted_fit} and
the full posterior in \autoref{fig:ted_corner}
(\autoref{sec:corners}).  What is remarkable is that the model can
provide a good simultaneous fit to the radial distributions of CO and
the dust continuum, as well as the integrated profiles.  We find that
for the best-fit model, the intrinsic gas distribution has an
$r_{\rm eff, mol} \approx 4$\,kpc,\footnote{Note the input radius is
  not the effective radius, which is instead computed from the gas
  distribution a posteriori (see \autoref{fig:ted_profiles}).} and an
$n\approx 1$.  The implied $r_{\rm eff, CO} \approx 4$\,kpc is
consistent with \cite{Hodge2012}, and the
$r_{\rm eff, 220\,\micron} \approx 2.5$\,kpc is in good agreement with
the S\'{e}rsic fits (noting the differences in $n$, see
\autoref{sec:dust-emission}).  We derive a total molecular gas mass
from the TED model of
$\Mmol = 2.9_{-0.3}^{+0.4} \times 10^{11}$\,\Msun with an
$\aco = 2.8_{-0.3}^{+0.5}$.  These estimates are very consistent with
the estimates from the unresolved data but with smaller error (cf.\
\autoref{sec:turb}).

A key result of the resolved modeling is that to first order the CO
and the dust emission can be described by a single underlying gas
density distribution, with a constant CO abundance and \gdr.  This
implies the different observed profiles for the dust and CO are not
necessarily due to intrinsically different spatial distributions of
the gas and dust, but rather due to radiative transfer effects.  We
further elaborate on this point in \autoref{sec:interpr-radi-prof}.

To second order, there are some discrepancies between the model and
the observations, most notably that the high-fidelity NOEMA profile at
261\,GHz rises more sharply to the center than the earlier PdBI
340\,GHz profile.  While the model takes into account beam
convolution, this cannot be fully described by the model
self-consistently under the assumed abundances and temperature
gradient, despite the model taking into account optical depth effects.
This could also be a limitation of the observations and/or
self-calibration, noting that both the PdBI and CO data have
significantly lower S/N than the high-fidelty NOEMA observations.

The assumption of a radial temperature gradient is motivated by both
theoretical considerations and observations of nearby galaxies
\citep[e.g.,][]{Galametz2012, Casasola2017}, and also observed in
simulated galaxies \citep{Cochrane2019}.  However, it does not have to
be linear and it may be that the true temperature gradient is steeper
in the central regions compared to the (outskirts of) the
disk. Indeed, we find that the best-fit model prefers a relatively
steep gradient, that artificially flattens in the outskirts because
the temperatures approach the temperature floor.  The relatively steep
gradient found in GN20 can therefore be interpreted an average
gradient, driven by the central regions.  For comparison, we also
tested a model without a temperature gradient (fixing $dT/dr = 0$).
However, we find this model provides a much poorer fit, as it fails to
match both the profile shapes and the integrated fluxes
simultaneously.

\section{Discussion} \label{sec:discussion}

\subsection{GN20: an extended, interaction-driven starburst}
\label{sec:nature-extend-starb}

The sensitive NOEMA imaging detects the dust continuum emission out to
large radii in GN20 and shows the dust extends over the full extent of
the stellar and molecular gas disk. The lower surface brightness dust
in the outskirts was not clearly detected in the earlier PdBI
observations \citep{Hodge2015} and emphasizes the importance of
sufficient sensitivity when mapping the dust emission in galaxies.
About 40\% of the total dust luminosity comes from radii outside of
the brightest nuclear region, implying that the starburst is very
extended, stretching over the entire disk of GN20.

Despite the disk-like ionized- and molecular gas kinematics
\citep{Hodge2012, Bik2024, Ubler2024}, there are several indications
that GN20 has undergone a recent interaction or merger.  Beyond its
location in an overdense protocluster environment \citep{Daddi2009},
these include an offset nucleus and a potentially disturbed morphology
with arm or tidal feature in the north \citep{Colina2023}, and the
presence of several clumps and nearby companions seen in \Halpha\
emission \citep[][]{Ubler2024}.  However, as pointed out in those
works, these features could also be caused by strongly variable
extinction in the dust-rich starburst.  The NIRCam and MIRI imaging
now reveals more clearly what is an arm or tidal feature to the north.
The arm or tidal feature stretches in the direction of the nearby
companion that was identified by its \Halpha\ emission
\citep{Ubler2024} and is now detected with NIRCam.  At the same time,
there is no elevated dust emission detected by NOEMA in the `gap' in
the stellar emission (rather, the ionized gas maps suggest the
extinction is lower in these regions).  This implies that the
morphological disturbances, including the arm- or tail-like feature,
that are seen in the MIRI and long-wavelength NIRCam imaging are
likely real and not an effect of extinction.  Taken together, these
clues provide increasingly strong evidence for the scenario in which
GN20 has undergone a recent interaction or merger, that is likely what
triggered or enhanced the very intense starburst in the system.

\subsection{Coincident gas and dust throughout the starburst: the
  importance of accounting for radiative transfer}
\label{sec:interpr-radi-prof}

\begin{figure*}[t]
  \centering
  \includegraphics[height=6.5cm]{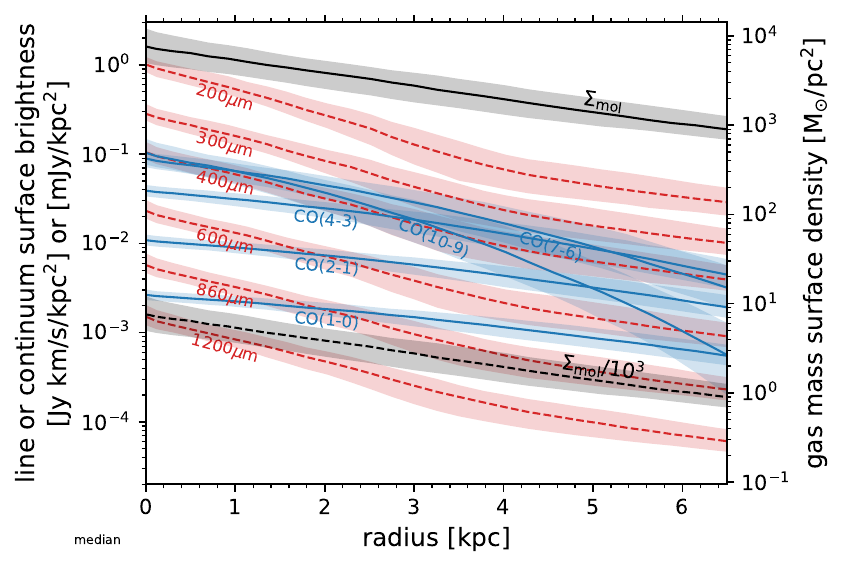}
  \includegraphics[height=6.5cm]{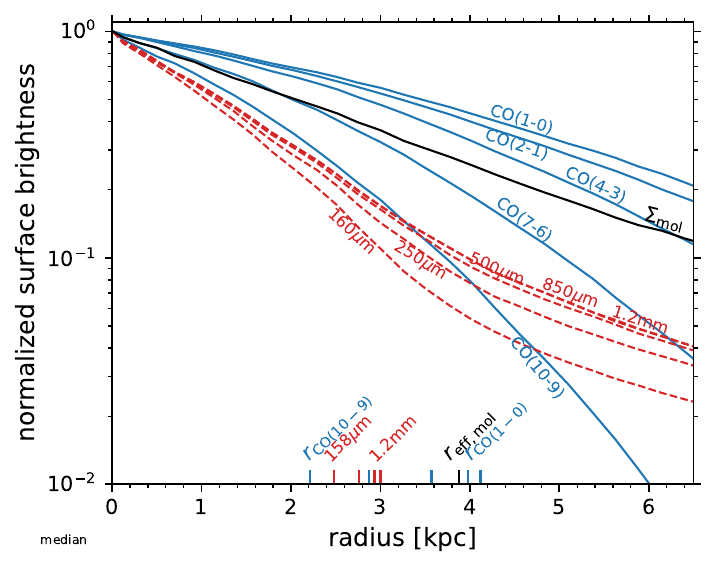}
  \caption{\label{fig:ted_profiles} \textbf{Left:} Median CO (blue)
    and dust (red) continuum surface brightness profiles from best-fit
    model together with the underlying gas mass surface density
    (black). \textbf{Right:} Same figure as on the left but now
    normalized to the values in the center. The measured effective
    (half-light) radii are indicated on the x-axis (with the lowest
    and highest-frequency line and transition
    labeled).}
\end{figure*}

\begin{figure}[t]
  \centering
  \includegraphics[width=\columnwidth]{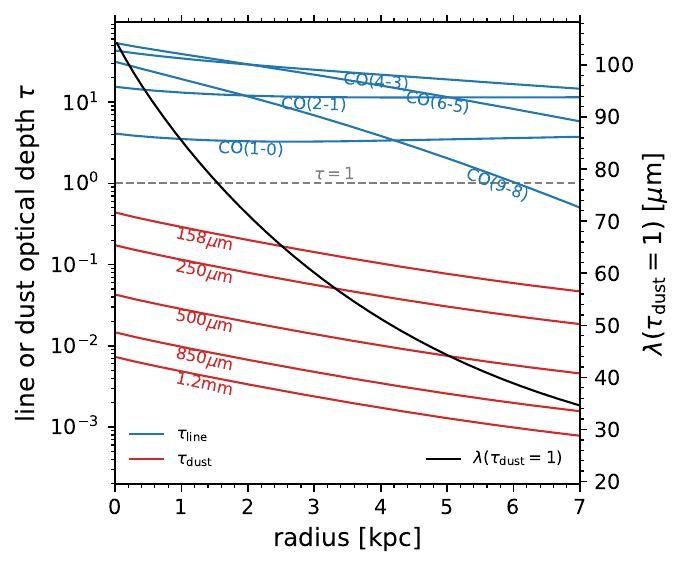}
  \caption{\label{fig:ted_tautex_prof} The radial line and continuum
    optical depths (blue and red lines, left axis) and rest-frame
    wavelengths where the optical depth reaches unity from the TED
    model (black line, right axis).  The low-$J$ CO lines are
    optically thick throughout.  The dust is mostly optically thin at
    rest-frame wavelengths longer than 100\,\micron, but becomes
    optically thick at those wavelengths in the center.}
\end{figure}

An important result from the resolved radiative transfer modeling in
\autoref{sec:ted} is that, to first order, the observed radial
profiles of the CO and the dust emission can be accurately described
using a single underlying gas density distribution with constant
\COHH\ and \gdr.  This implies that the different observed profiles
for the dust and CO are not due to intrinsically different spatial
distributions of the gas and dust, but rather due to radiative
transfer effects.  Such behavior has been suggested before from the
average CO and dust continuum profiles in stacked samples of SMGs by
\cite{Calistro2018}.

Specifically, the results imply that the smaller `size' of the dust
distribution compared to the gas is not due to the dust being actually
more concentrated in the center of the galaxy, but can simply be due
to radiative transfer.  To explain this, we look at the radial
profiles of the molecular gas mass surface density, as well as
selected CO transitions and the dust at different wavelengths for the
best-fit model of GN20 in \autoref{fig:ted_profiles}.  We also refer
the reader to the equations of radiative transfer given in
\autoref{sec:tuner}.

In \autoref{fig:ted_profiles}, the molecular gas mass surface
distribution is shown by the black line.  Looking at the shape of the
surface brightness profiles, the low-$J$ CO lines trace the intrinsic
shape of the molecular gas surface density profile closely in the
outskirts, slightly flattening towards the center, and overall better
than the 1.2\,mm dust continuum that shows a steeper profile.

This behavior can be explained through radiative transfer as follows:
while the low-$J$ CO line emission is optically thick throughout
($\tau_{\nu_L} \geq 1$), it is sub-thermally excited in the outskirts
($\Texc < \Tkin$).  Moving radially inward, the increasing emission is
due to the increasing CO excitation as the gas surface density
increases, until the CO is both optically thick and thermalized
($\Texc = \Tkin$; and thus saturates)\footnote{Note it is a
  misconception that the optically thick $^{12}$CO emission is not
  sensitive to the gas mass surface density (or column density); it
  still is as long as it is subthermally excited, as \Texc\ still
  increases with higher density (cf.\ \autoref{eq:rt_cont}).} at which
point it can only increases further due to the increasing kinetic
temperature (via the radial temperature gradient).  In contrast, the
(longer-wavelength) dust is optically thin throughout most of the
galaxy ($\tau_{\rm \nu, dust} < 1$), and hence moving radially inward,
the emission becomes brighter towards the center both because of the
increasing column density (increasing $\tau_{\rm \nu, dust}$) and the
increasing dust temperature \Tdust, and thus follows a steeper light
profile (even if it becomes optically thick in the
center). %
As such, even though the gas and dust share the same underlying mass
profile and temperature gradient, the emitted profile for
the CO and dust continuum is different.

From \autoref{fig:ted_profiles} and the above it also follows that the
low-$J$ CO traces the shape of the underlying \HH-mass profile more
closely than the mm-continuum, albeit not perfectly if a single
conversion factor (and excitation correction for the $J_{\rm up}>1$
lines) is used.

It should go without saying that the detailed behavior seen in
\autoref{fig:ted_profiles} is specific to GN20, and variations in this
picture are expected for different types of galaxies.  However the
qualitative behavior of the CO and dust---driven by the former being
optically thick in general (at least in the low-$J$ lines) while the
latter is optically thin---should hold more generally.

The broader implication is that that careful consideration of
radiative transfer effects is important when interpreting observed
profiles from gas and dust tracers.  A particular example is when
comparing gas and dust sizes.  Note that while the \emph{observed}
effective radii of the CO and dust emission are significantly
different, the \emph{intrinsic} effective radii of the gas and dust
distributions in the model are exactly the same.  A face-to-face
comparison of the effective radii or other measures of the observed
profiles between different gas and dust emission to measure their size
or extent is therefore misleading.

\subsection{Deceptively optically thick dust in GN20?}
\label{sec:opt-thin}

The median dust optical depths predicted from the TED model are shown
in \autoref{fig:ted_tautex_prof}.  The dust is globally optically thin
at wavelengths longer than $\approx$100\,\micron.  At the effective
radius, the wavelength where the dust becomes optically thick is
$\approx 70$\,\micron, very consistent with average value of
$\approx70$\,\micron\ from the unresolved TUNER model.\footnote{The
  dust optical depth $\tau = \Sigma_d \kappa_d$, where $\Sigma_d$ is
  the dust mass surface density and $\kappa_d$ the dust opacity, and
  the wavelength where the dust becomes optically thick is thus
  $\lambda(\tau=1) = (\Sigma_{d} \kappa_{d, 0})^{1/\beta} \lambda_0$.}

The dust optical depths from the model are significantly lower than
previously reported by \cite{Cortzen2020}, found that the global
wavelength where the dust gets optically thick is
$170 \pm 23$\,\micron\ (for a very similar dust opacity of
0.43\,cm$^{2}$\,g$^{-1}$ from \citealt{Li2001}) resulting in a higher
\Tdust. The high optical depth found in \cite{Cortzen2020} corresponds
to an radius of $\approx$1.2\,kpc, which they noted to be broadly
consistent with the effective radius of the PdBI data from
\cite{Hodge2015}.  Looking in detail, however, the deconvolved FWHM
size measured by \cite[][]{Hodge2015} in fact corresponds to a radius
of $\approx 2$\,kpc (which can also be read-off directly from
\autoref{fig:radial-profiles}).  This implies that the actual dust
column is lower and the wavelength where the dust is optically thick
(using their values) comes closer to $\sim$100 micron.  Note that
there is no physical mechanism that requires
$\Tdust = T_{\rm exc, \CI}$ and, moreover, the derivation of the
latter requires several assumptions (notably that \CI\ is in LTE,
which are not a given).

We stress that the exact profiles and opacities that result from the
model are sensitive to the model assumptions as well as systematics in
the datasets and should not be overinterpreted.  However, the
qualitative results explaining the observed radial profiles and the
lower dust optical depth stand.  A simple test distributing the total
dust mass derived from the models over the NOEMA 1.1\,mm map (that is,
assuming this is a dust-mass map, which is an upper limit on the
actual central opacity in the case of dust temperature gradients)
shows $\lambda(\tau=1)\approx140$\,\micron\ in the central pixel,
confirming that it is hard to conceive that global average optical
depth is that high.

The picture that emerges is as follows: while GN20 indeed has a very
large dust mass, it also hosts a substantial disk, and the dust mass
distribution in GN20 is quite extended.  As a result, while global
dust columns are large, they are also not exceptionally large, as
seen, for example, on small scales in the nuclei of some galaxies
(e.g., Arp 220; \citealt{Scoville2017}).  The dust column densities
and \HH-column densities predicted by the model are still substantial,
of order $N_{\HH} \approx 10^{23}$\,cm$^{-2}$.  These imply an
extinction of $A_V \approx 23$ magnitudes \citep{Bohlin1978,
  Guver2009}, consistent with values derived from \Paa\
\citep{Bik2024}, and ample to completely obscure GN20 in the
rest-frame optical.

\section{Conclusions} \label{sec:conclusions}

We present sensitive and high resolution NOEMA imaging at
0.9--1.6\,kpc resolution (0\farcs13--0\farcs23) of the extended dust
continuum emission at 1.1\,mm (rest-frame 220\,\micron) in the
$z=4.055$ dusty star-forming galaxy GN20.  Together with JWST/MIRI and
NIRCam imaging, that now resolves the (previously-obscured) morphology
in the rest-frame optical, near- and mid-infrared, we study the
distribution of gas, dust and stars in this prototypical dusty
starburst galaxy.

The main conclusions of this paper are as follows:

\begin{itemize}
\item The 1.1\,mm dust emission is extended and now detected over the
  full stellar and molecular gas disk of GN20, sharing a common
  center.  The dust emission stretches $\approx$14\,kpc\ in diameter
  on its longest axis and is centrally asymmetric and clumpy, with
  $r_e \approx 2.5$\,kpc and $b/a=0.5$, being brightest in the
  strongly obscured part of the galaxy.  Only 60\% of the total dust
  emission arises from the central $3.5\times1.5$\,kpc
  ($0\farcs5$--$0\farcs2$) and one-third from the nucleus and the
  prominent clump to the south, implying the starburst is very
  extended.

\item NIRCam and MIRI imaging in the rest-frame optical and
  near-infrared reveal a clear arm or tidal tail feature in the disk
  of GN20, the outskirts of the latter coincide with the
  least-obscured western part of the galaxy detected in the
  UV/optical.  The dust emission in the center is asymmetric and
  bar-shaped, with a bright dust clump at the tip.  There is a
  prominent gap in the stellar emission between the arm and the
  nucleus, that is likely a real depression and not due to extinction,
  supported by the lack of elevated dust emission.  Together with the
  offset nucleus, the presence of nearby companions and the location
  in a protocluster, this strongly supports GN20 having experienced a
  recent interaction or merger, which has likely invigorated the
  starburst.

\item We perform self-consistent radiative transfer modeling of the CO
  and dust-continuum emission using the Turbulent Non-Equilibrium
  Radiative Transfer (TUNER) model to derive the molecular gas and
  dust masses of the system.  Interestingly, under the assumed
  \cite{Draine2014} dust opacity, we find a low (super-solar)
  gas-to-dust ratio of $\gdr \approx 50$ (and a correspondingly high
  dust mass $\approx5.8\times10^{9}$\,\Msun) with an approximately
  solar CO abundance ($\COHH \approx 10^{-4}$).  This may indicate
  significantly different dust properties in the dust-rich starburst.

\item The radial surface brightness profiles of the CO and near-IR
  stellar emission are similar ($r_e\approx4$\,kpc), suggesting a
  roughly constant gas fraction across the GN20 disk.  In contrast,
  the dust emission appears significantly more concentrated
  ($r_e\approx2.5$\,kpc).  By coupling the TUNER model to a S\'{e}rsic
  column density distribution with radially decreasing linear
  temperature gradient, we show that this does not mean that the dust
  is intrinsically more compact.  Instead, the profiles can be
  naturally explained by differences in the radiative transfer between
  the (mostly) optically thin dust and optically thick CO.  The
  observations are consistent with the gas and dust mass being
  similarly distributed throughout the starburst.

\item From the combined modeling of the unresolved CO and dust SED
  together with the resolved radial profiles, we derive a total
  molecular gas mass in GN20 of
  $\Mmol \approx 2.9^{+0.4}_{-0.3} \times 10^{11}$\,\Msun with an
  $\aco = 2.8^{+0.5}_{-0.3}$.  Given the dynamical mass of
  $\approx 5\times10^{11}$\,\Msun, this implies GN20 is predominately
  molecular gas rich.

\item We find the dust emission is globally optically thick at
  wavelengths $\leq100$\,\micron, lower than previously reported by
  \cite{Cortzen2020}, but consistent with the extended dust
  distribution and sufficiently high dust column densities to
  completely obscure GN20 in the rest-frame optical.

\end{itemize}

This study demonstrates the incredible combined power of JWST and
(sub)mm interferometers like NOEMA and ALMA to resolve the stellar,
gas and dust structure of dusty star-forming galaxies in the early
universe.  However, our work shows that it remains imperative to
carefully consider radiative transfer effects when comparing molecular
gas and dust distributions from different tracers.  Future studies
leveraging deep and high-resolution (sub)mm observations, such as the
ALMA Large Programs CONDOR (2024.1.00100.L) and HIDING in the HUDF
(2025.1.01377.L), will be able to resolve the gas, dust and stars in
more typical early galaxies.

\begin{acknowledgments}
  We thank the referee for a constructive report that helped improve
  the paper.  We are grateful to Luca Costantin for providing the
  NIRCam imaging.  We are also grateful to Hannah \"{U}bler for
  providing the \Ha\ map.  L.A.B. acknowledges support from the Dutch
  Research Council (NWO) under grant VI.Veni.242.055
  (\url{https://doi.org/10.61686/LAJVP77714}).  L.A.B. and
  J.H. acknowledge support from the ERC Consolidator grant 101088676
  ("VOYAJ").  L.C. acknowledges support by grant PIB2021-127718NB-100
  from the Spanish Ministry of Science and Innovation/State Agency of
  Research MCIN/AEI/10.13039/501100011033 and by “ERDF A way of making
  Europe. A.C.G.  acknowledges support by JWST contract
  B0215/JWST-GO-02926.  A.B. and G.O. acknowledges support from the
  Swedish National Space Administration (SNSA).  This work is based on
  observations carried out under project number W22DU with the IRAM
  NOEMA Interferometer.  IRAM is supported by INSU/CNRS (France), MPG
  (Germany) and IGN (Spain).  Some of the data presented in this
  article were obtained from the Mikulski Archive for Space Telescopes
  (MAST) at the Space Telescope Science Institute. The specific
  observations analyzed can be accessed via
  \dataset[doi:10.17909/mrb8-x762]{https://doi.org/10.17909/mrb8-x762}.
\end{acknowledgments}

\vspace{5mm}
\facilities{NOEMA (IRAM)}

\software{\textsc{topcat} \citep{taylor2005}, \textsc{gnuastro}
  \citep{akhlaghi2015}, \textsc{ipython} \citep{perez2007},
  \textsc{numpy} \citep{numpy2020}, \textsc{scipy} \citep{scipy2020},
  \textsc{matplotlib} \citep{hunter2007}, \textsc{astropy}
  \citep{TheAstropyCollaboration2022}, \textsc{photutils}
  \citep{photutils160}, \textsc{spectralcube} \citep{spectralcube},
  \textsc{interferopy} \citep{interferopy}.}

\bibliography{library}{}

\begin{thebibliography}{}
\expandafter\ifx\csname natexlab\endcsname\relax\def\natexlab#1{#1}\fi
\providecommand{\url}[1]{\href{#1}{#1}}
\providecommand{\dodoi}[1]{doi:~\href{http://doi.org/#1}{\nolinkurl{#1}}}
\providecommand{\doeprint}[1]{\href{http://ascl.net/#1}{\nolinkurl{http://ascl.net/#1}}}
\providecommand{\doarXiv}[1]{\href{https://arxiv.org/abs/#1}{\nolinkurl{https://arxiv.org/abs/#1}}}

\bibitem[{Akhlaghi \& Ichikawa(2015)}]{akhlaghi2015}
Akhlaghi, M., \& Ichikawa, T. 2015, \apjs, 220, 1,
  \dodoi{10.1088/0067-0049/220/1/1}

\bibitem[{{\'{A}}lvarez-M{\'{a}}rquez
  {et~al.}(2023){\'{A}}lvarez-M{\'{a}}rquez, {Crespo G{\'{o}}mez}, Colina,
  Neeleman, Walter, Labiano, P{\'{e}}rez-Gonz{\'{a}}lez, Bik,
  Noorgaard-Nielsen, Ostlin, Wright, Alonso-Herrero, Azollini, Caputi, Eckart,
  {Le F{\`{e}}vre}, Garc{\'{i}}a-Mar{\'{i}}n, Greve, Hjorth, Ilbert, Kendrew,
  Pye, Tikkanen, Topinka, {Van Der Werf}, Ward, {Van Dishoeck}, G{\"{u}}del,
  Henning, Lagage, Ray, \& Waelkens}]{Alvarez-Marquez2023}
{\'{A}}lvarez-M{\'{a}}rquez, J., {Crespo G{\'{o}}mez}, A., Colina, L., {et~al.}
  2023, \aap, 671, \dodoi{10.1051/0004-6361/202245400}

\bibitem[{Asplund {et~al.}(2009)Asplund, Grevesse, Sauval, \&
  Scott}]{Asplund2009}
Asplund, M., Grevesse, N., Sauval, A.~J., \& Scott, P. 2009, \araa, 47, 481,
  \dodoi{10.1146/annurev.astro.46.060407.145222}

\bibitem[{Bagley {et~al.}(2023)Bagley, Finkelstein, Koekemoer, Ferguson,
  {Arrabal Haro}, Dickinson, Kartaltepe, Papovich, P{\'{e}}rez-Gonz{\'{a}}lez,
  Pirzkal, Somerville, Willmer, Yang, Yung, Fontana, Grazian, Grogin,
  Hirschmann, Kewley, Kirkpatrick, Kocevski, Lotz, Medrano, Morales,
  Pentericci, Ravindranath, Trump, Wilkins, Calabr{\`{o}}, Cooper, Costantin,
  de~la Vega, Hilbert, Hutchison, Larson, Lucas, McGrath, Ryan, Wang, \&
  Wuyts}]{Bagley2023}
Bagley, M.~B., Finkelstein, S.~L., Koekemoer, A.~M., {et~al.} 2023, \apjl, 946,
  L12, \dodoi{10.3847/2041-8213/acbb08}

\bibitem[{Bik {et~al.}(2024)Bik, Alvarez-Marquez, Colina, {Crespo Gomez},
  Peissker, Walter, Boogaard, AOstlin, Greve, Wright, Alonso-Herrero, Caputi,
  Costantin, Eckart, Gillman, Hjorth, Iani, Jermann, Labiano, Langeroodi,
  Melinder, Perez-Gonzalez, Pye, Rinaldi, Tikkanen, {Van Der Werf}, Gudel,
  Henning, Lagage, Ray, \& {Van Dishoeck}}]{Bik2024}
Bik, A., Alvarez-Marquez, J., Colina, L., {et~al.} 2024, \aap, 686, 1,
  \dodoi{10.1051/0004-6361/202348845}

\bibitem[{Bohlin \& Drake(1978)}]{Bohlin1978}
Bohlin, S., \& Drake. 1978, ApJ, 132

\bibitem[{Boogaard {et~al.}(2021)Boogaard, Meyer, \& Novak}]{interferopy}
Boogaard, L., Meyer, R.~A., \& Novak, M. 2021, {Interferopy: analysing
  datacubes from radio-to-submm observations}, \dodoi{10.5281/ZENODO.5775603}

\bibitem[{Boogaard {et~al.}(2024)Boogaard, Gillman, Melinder, Walter, Colina,
  {\"{O}}stlin, Caputi, Iani, P{\'{e}}rez-Gonz{\'{a}}lez, van~der Werf, Greve,
  Wright, Alonso-Herrero, {\'{A}}lvarez-M{\'{a}}rquez, Annunziatella, Bik,
  Bosman, Costantin, {Crespo G{\'{o}}mez}, Dicken, Eckart, Hjorth, Jermann,
  Labiano, Langeroodi, Meyer, Moutard, Pei{\ss}ker, Pye, Rinaldi, Tikkanen,
  Topinka, \& Henning}]{Boogaard2024}
Boogaard, L.~A., Gillman, S., Melinder, J., {et~al.} 2024, \apj, 969, 27,
  \dodoi{10.3847/1538-4357/ad43e5}

\bibitem[{Bradley {et~al.}(2022)Bradley, Sipőcz, Robitaille, Tollerud,
  Vin{\'{i}}cius, Deil, Barbary, Wilson, Busko, Donath, G{\"{u}}nther, Cara,
  Lim, Me{\ss}linger, Conseil, Bostroem, Droettboom, Bray, Bratholm, Barentsen,
  Craig, Ginsburg, Rathi, Pascual, Perren, Georgiev, de~Val-Borro, Kerzendorf,
  Bach, \& Quint}]{photutils160}
Bradley, L., Sipőcz, B., Robitaille, T., {et~al.} 2022, astropy/photutils:
  1.6.0,  Zenodo, \dodoi{10.5281/zenodo.7419741}

\bibitem[{{Calistro Rivera} {et~al.}(2018){Calistro Rivera}, Hodge, Smail,
  Swinbank, Weiss, Wardlow, Walter, Rybak, Chen, Brandt, Coppin, da~Cunha,
  Dannerbauer, Greve, Karim, Knudsen, Schinnerer, Simpson, Venemans, \& van~der
  Werf}]{Calistro2018}
{Calistro Rivera}, G., Hodge, J.~A., Smail, I., {et~al.} 2018, \apj, 863, 56,
  \dodoi{10.3847/1538-4357/aacffa}

\bibitem[{Carilli {et~al.}(2011)Carilli, Hodge, Walter, Riechers, Daddi,
  Dannerbauer, \& Morrison}]{Carilli2011}
Carilli, C.~L., Hodge, J., Walter, F., {et~al.} 2011, \apj, 739, L33,
  \dodoi{10.1088/2041-8205/739/1/L33}

\bibitem[{Carilli {et~al.}(2010)Carilli, Daddi, Riechers, Walter, Weiss,
  Dannerbauer, Morrison, Wagg, Dav{\'{e}}, Elbaz, Stern, Dickinson, Krips, \&
  Aravena}]{Carilli2010}
Carilli, C.~L., Daddi, E., Riechers, D., {et~al.} 2010, \apj, 714, 1407,
  \dodoi{10.1088/0004-637X/714/2/1407}

\bibitem[{Casasola {et~al.}(2017)Casasola, Cassar{\`{a}}, Bianchi, Verstocken,
  Xilouris, Magrini, Smith, {De Looze}, Galametz, Madden, Baes, Clark, Davies,
  {De Vis}, Evans, Fritz, Galliano, Jones, Mosenkov, Viaene, \&
  Ysard}]{Casasola2017}
Casasola, V., Cassar{\`{a}}, L.~P., Bianchi, S., {et~al.} 2017, \aap, 605,
  \dodoi{10.1051/0004-6361/201731020}

\bibitem[{Casey {et~al.}(2014)Casey, Narayanan, \& Cooray}]{Casey2014}
Casey, C.~M., Narayanan, D., \& Cooray, A. 2014, Phys. Rep., 541, 45,
  \dodoi{10.1016/j.physrep.2014.02.009}

\bibitem[{Casey {et~al.}(2009)Casey, Chapman, Daddi, Dannerbauer, Pope, Scott,
  Bertoldi, Beswick, Blain, Cox, Genzel, Greve, Ivison, Muxlow, Neri, Omont,
  Smail, \& Tacconi}]{Casey2009}
Casey, C.~M., Chapman, S.~C., Daddi, E., {et~al.} 2009, \mnras, 400, 670,
  \dodoi{10.1111/j.1365-2966.2009.15517.x}

\bibitem[{Chen {et~al.}(2017)Chen, Hodge, Smail, Swinbank, Walter, Simpson,
  Rivera, Bertoldi, Brandt, Chapman, da~Cunha, Dannerbauer, Breuck, Harrison,
  Ivison, Karim, Knudsen, Wardlow, Wei{\ss}, \& van~der Werf}]{Chen2017}
Chen, C.-C., Hodge, J.~A., Smail, I., {et~al.} 2017, \apj, 846, 108,
  \dodoi{10.3847/1538-4357/aa863a}

\bibitem[{Cheng {et~al.}(2022)Cheng, Yan, Huang, Willmer, Ma, \&
  Orellana-Gonz{\'{a}}lez}]{Cheng2022}
Cheng, C., Yan, H., Huang, J.-S., {et~al.} 2022, \apjl, 936, L19,
  \dodoi{10.3847/2041-8213/ac8d08}

\bibitem[{Cochrane {et~al.}(2019)Cochrane, Hayward, Angl{\'{e}}s-Alc{\'{a}}zar,
  Lotz, Parsotan, Ma, Kere{\v{s}}, Feldmann, Faucher-Gigu{\`{e}}re, \&
  Hopkins}]{Cochrane2019}
Cochrane, R.~K., Hayward, C.~C., Angl{\'{e}}s-Alc{\'{a}}zar, D., {et~al.} 2019,
  \mnras, 488, 1779, \dodoi{10.1093/mnras/stz1736}

\bibitem[{Colina {et~al.}(2023)Colina, {Crespo G{\'{o}}mez},
  {\'{A}}lvarez-M{\'{a}}rquez, Bik, Walter, Boogaard, Labiano, Peissker,
  P{\'{e}}rez-Gonz{\'{a}}lez, {\"{O}}stlin, Greve, N{\o}rgaard-Nielsen, Wright,
  Alonso-Herrero, Azollini, Caputi, Dicken, Garc{\'{i}}a-Mar{\'{i}}n, Hjorth,
  Ilbert, Kendrew, Pye, Tikkanen, {Van Der Werf}, Costantin, Iani, Gillman,
  Jermann, Langeroodi, Moutard, Rinaldi, Topinka, {Van Dishoeck}, G{\"{u}}del,
  Henning, Lagage, Ray, \& Vandenbussche}]{Colina2023}
Colina, L., {Crespo G{\'{o}}mez}, A., {\'{A}}lvarez-M{\'{a}}rquez, J., {et~al.}
  2023, \aap, 673, 1, \dodoi{10.1051/0004-6361/202346535}

\bibitem[{Cortzen {et~al.}(2020)Cortzen, Magdis, Valentino, Daddi, Liu,
  Rigopoulou, Sargent, Riechers, Cormier, Hodge, Walter, Elbaz,
  B{\'{e}}thermin, Greve, Kokorev, \& Toft}]{Cortzen2020}
Cortzen, I., Magdis, G.~E., Valentino, F., {et~al.} 2020, \aap, 634, 1,
  \dodoi{10.1051/0004-6361/201937217}

\bibitem[{{Crespo G{\'{o}}mez} {et~al.}(2024){Crespo G{\'{o}}mez}, Colina,
  {\'{A}}lvarez-M{\'{a}}rquez, Bik, Boogaard, {\"{O}}stlin, Pei{\ss}ker,
  Walter, Labiano, P{\'{e}}rez-Gonz{\'{a}}lez, Greve, Wright, Alonso-Herrero,
  Caputi, Costantin, Eckart, Garc{\'{i}}a-Mar{\'{i}}n, Gillman, Hjorth, Iani,
  Langeroodi, Pye, Rinaldi, Tikkanen, van~der Werf, Lagage, \& van
  Dishoeck}]{Crespo2024}
{Crespo G{\'{o}}mez}, A., Colina, L., {\'{A}}lvarez-M{\'{a}}rquez, J., {et~al.}
  2024, 1.
\newblock \doarXiv{2402.18672}

\bibitem[{{Da Cunha} {et~al.}(2015){Da Cunha}, Walter, Smail, Swinbank,
  Simpson, Decarli, Hodge, Weiss, Werf, Bertoldi, Chapman, Cox, Danielson,
  Dannerbauer, Greve, Ivison, Karim, \& Thomson}]{DaCunha2015}
{Da Cunha}, E., Walter, F., Smail, I.~R., {et~al.} 2015, \apj, 806, 110,
  \dodoi{10.1088/0004-637X/806/1/110}

\bibitem[{Daddi {et~al.}(2009)Daddi, Dannerbauer, Stern, Dickinson, Morrison,
  Elbaz, Giavalisco, Mancini, Pope, \& Spinrad}]{Daddi2009}
Daddi, E., Dannerbauer, H., Stern, D., {et~al.} 2009, \apj, 694, 1517,
  \dodoi{10.1088/0004-637X/694/2/1517}

\bibitem[{Daddi {et~al.}(2015)Daddi, Dannerbauer, Liu, Aravena, Bournaud,
  Walter, Riechers, Magdis, Sargent, B{\'{e}}thermin, Carilli, Cibinel,
  Dickinson, Elbaz, Gao, Gobat, Hodge, \& Krips}]{Daddi2015}
Daddi, E., Dannerbauer, H., Liu, D., {et~al.} 2015, \aap, 577, A46,
  \dodoi{10.1051/0004-6361/201425043}

\bibitem[{Draine {et~al.}(2014)Draine, Aniano, Krause, Groves, Sandstrom,
  Braun, Leroy, Klaas, Linz, Rix, Schinnerer, Schmiedeke, \&
  Walter}]{Draine2014}
Draine, B.~T., Aniano, G., Krause, O., {et~al.} 2014, \apj, 780, 172,
  \dodoi{10.1088/0004-637X/780/2/172}

\bibitem[{Dunne {et~al.}(2000)Dunne, Eales, Edmunds, Ivison, Alexander, \&
  Clements}]{Dunne2000}
Dunne, L., Eales, S., Edmunds, M., {et~al.} 2000, \mnras, 315, 115,
  \dodoi{10.1046/j.1365-8711.2000.03386.x}

\bibitem[{Foreman-Mackey(2016)}]{Foreman-Mackey2016}
Foreman-Mackey, D. 2016, J. Open Source Softw., 1, 1,
  \dodoi{10.21105/joss.00024}

\bibitem[{Foreman-Mackey {et~al.}(2013)Foreman-Mackey, Hogg, Lang, \&
  Goodman}]{Foreman-Mackey2013}
Foreman-Mackey, D., Hogg, D.~W., Lang, D., \& Goodman, J. 2013, \pasp, 125,
  306, \dodoi{10.1086/670067}

\bibitem[{Frerking {et~al.}(1982)Frerking, Langer, \& Wilson}]{Frerking1982}
Frerking, M.~A., Langer, W.~D., \& Wilson, R.~W. 1982, \apj, 262, 590,
  \dodoi{10.1086/160451}

\bibitem[{Galametz {et~al.}(2012)Galametz, Kennicutt, Albrecht, Aniano, Armus,
  Bertoldi, Calzetti, Crocker, Croxall, Dale, {Donovan Meyer}, Draine,
  Engelbracht, Hinz, Roussel, Skibba, Tabatabaei, Walter, Weiss, Wilson, \&
  Wolfire}]{Galametz2012}
Galametz, M., Kennicutt, R.~C., Albrecht, M., {et~al.} 2012, \mnras, 425, 763,
  \dodoi{10.1111/j.1365-2966.2012.21667.x}

\bibitem[{Gillman {et~al.}(2023)Gillman, Gullberg, Brammer, Vijayan, Lee,
  Bl{\'{a}}nquez, Brinch, Greve, Jermann, Jin, Kokorev, Liu, Magdis, Rizzo, \&
  Valentino}]{Gillman2023}
Gillman, S., Gullberg, B., Brammer, G., {et~al.} 2023, \aap, 676, A26,
  \dodoi{10.1051/0004-6361/202346531}

\bibitem[{Gillman {et~al.}(2024)Gillman, Smail, Gullberg, Swinbank, Vijayan,
  Lee, Brammer, Dudzevi{\v{c}}iut, Greve, Almaini, Brinch, Chapman, Chen,
  Ikarashi, Matsuda, Wang, Walter, \& {Van Der Werf}}]{Gillman2024}
Gillman, S., Smail, I., Gullberg, B., {et~al.} 2024, \aap, 691, 1,
  \dodoi{10.1051/0004-6361/202451006}

\bibitem[{Ginsburg {et~al.}(2019)Ginsburg, Koch, Robitaille, Beaumont,
  Adamginsburg, Sipőcz, ZuHone, Patra, Jones, Lim, Stern, Rosolowsky, Earl,
  de~Val-Borro, Jrobbfed, Shuokong, Kepley, Sokolov, Badger, Maret, Garrido,
  Booker, \& Tollerud}]{spectralcube}
Ginsburg, A., Koch, E., Robitaille, T., {et~al.} 2019,
  {radio-astro-tools/spectral-cube: Release v0.4.5},  Zenodo,
  \dodoi{10.5281/zenodo.3558614}

\bibitem[{Girard {et~al.}(2021)Girard, Fisher, Bolatto, Abraham, Bassett,
  Glazebrook, Herrera-Camus, Jim{\'{e}}nez, Lenki{\'{c}}, \&
  Obreschkow}]{Girard2021}
Girard, M., Fisher, D.~B., Bolatto, A.~D., {et~al.} 2021, \apj, 909, 12,
  \dodoi{10.3847/1538-4357/abd5b9}

\bibitem[{Goldsmith(2001)}]{Goldsmith2001}
Goldsmith, P.~F. 2001, \apj, 557, 736, \dodoi{10.1086/322255}

\bibitem[{Goodman \& Weare(2010)}]{Goodman2010}
Goodman, J., \& Weare, J. 2010, Commun. Appl. Math. Comput. Sci., 5, 65,
  \dodoi{10.2140/camcos.2010.5.65}

\bibitem[{Greve {et~al.}(2009)Greve, Papadopoulos, Gao, \& Radford}]{Greve2009}
Greve, T.~R., Papadopoulos, P.~P., Gao, Y., \& Radford, S.~J. 2009, \apj, 692,
  1432, \dodoi{10.1088/0004-637X/692/2/1432}

\bibitem[{Gullberg {et~al.}(2019)Gullberg, Smail, Swinbank,
  Dudzevi{\v{c}}iūtė, Stach, Thomson, Almaini, Chen, Conselice, Cooke,
  Farrah, Ivison, Maltby, Micha{\l}owski, Simpson, Scott, Wardlow, \&
  Weiss}]{Gullberg2019}
Gullberg, B., Smail, I., Swinbank, A.~M., {et~al.} 2019, \mnras, 490, 4956,
  \dodoi{10.1093/mnras/stz2835}

\bibitem[{G{\"{u}}ver \& {\"{O}}zel(2009)}]{Guver2009}
G{\"{u}}ver, T., \& {\"{O}}zel, F. 2009, \mnras, 400, 2050,
  \dodoi{10.1111/j.1365-2966.2009.15598.x}

\bibitem[{Harrington {et~al.}(2021)Harrington, Weiss, Yun, Magnelli, Sharon,
  Leung, Vishwas, Wang, Frayer, Jim{\'{e}}nez-Andrade, Liu, Garc{\'{i}}a,
  Romano-D{\'{i}}az, Frye, Jarugula, Bădescu, Berman, Dannerbauer,
  D{\'{i}}az-S{\'{a}}nchez, Grassitelli, Kamieneski, Kim, Kirkpatrick,
  Lowenthal, Messias, Puschnig, Stacey, Torne, \& Bertoldi}]{Harrington2021}
Harrington, K.~C., Weiss, A., Yun, M.~S., {et~al.} 2021, \apj, 908, 95,
  \dodoi{10.3847/1538-4357/abcc01}

\bibitem[{Harris {et~al.}(2020)Harris, Millman, van~der Walt, Gommers,
  Virtanen, Cournapeau, Wieser, Taylor, Berg, Smith, Kern, Picus, Hoyer, van
  Kerkwijk, Brett, Haldane, del R{\'{i}}o, Wiebe, Peterson,
  G{\'{e}}rard-Marchant, Sheppard, Reddy, Weckesser, Abbasi, Gohlke, \&
  Oliphant}]{numpy2020}
Harris, C.~R., Millman, K.~J., van~der Walt, S.~J., {et~al.} 2020, \nat, 585,
  357, \dodoi{10.1038/s41586-020-2649-2}

\bibitem[{Hodge {et~al.}(2012)Hodge, Carilli, Walter, de~Blok, Riechers, Daddi,
  \& Lentati}]{Hodge2012}
Hodge, J.~A., Carilli, C.~L., Walter, F., {et~al.} 2012, \apj, 760, 11,
  \dodoi{10.1088/0004-637X/760/1/11}

\bibitem[{Hodge \& da~Cunha(2020)}]{Hodge2020}
Hodge, J.~A., \& da~Cunha, E. 2020, R. Soc. Open Sci., 7, 200556,
  \dodoi{10.1098/rsos.200556}

\bibitem[{Hodge {et~al.}(2015)Hodge, Riechers, Decarli, Walter, Carilli, Daddi,
  \& Dannerbauer}]{Hodge2015}
Hodge, J.~A., Riechers, D., Decarli, R., {et~al.} 2015, \apj, 798, L18,
  \dodoi{10.1088/2041-8205/798/1/L18}

\bibitem[{Hodge {et~al.}(2016)Hodge, Swinbank, Simpson, Smail, Walter,
  Alexander, Bertoldi, Biggs, Brandt, Chapman, Chen, Coppin, Cox, Edge, Greve,
  Ivison, Karim, Knudsen, Menten, Rix, Schinnerer, Wardlow, Weiss, van~der
  Werf, Dannerbauer, Edge, Greve, Ivison, Karim, Knudsen, Menten, Rix,
  Schinnerer, Wardlow, Weiss, van~der Werf, {Agertz O.}, B., et~al {Aravena M.,
  Hodge J. A.}, et~al {Barcos-Mu{\~{n}}oz L., Leroy A. K.}, et~al {Barro G.,
  Kriek M.}, {Blain A. W.}, C., et~al {Blain A. W., Smail I.}, et~al {Bolatto
  A. D., Warren S. R.}, et~al {Bothwell M. S., Smail I.}, et~al {Bothwell M.
  S., Chapman S. C.}, F., et~al {Bournaud F., Chapon D.}, et~al {Bournaud F.,
  Juneau S.}, et~al {Bournaud F., Perret V.}, et~al {Cappellari M., Emsellem
  E.}, et~al {Cappellari M., McDermid R. M.}, et~al {Carilli C. L., Daddi E.},
  F., Walter, {Casey C. M.}, A., {Chapin E. L.}, I., {Chapman S. C., Blain A.
  W.}, J., {Chapman S. C., Helou G.}, A., et~al {Chapman S. C., Smail I.},
  {Chapman S. C., Windhorst R., Odewahn S.}, C., et~al {Chen C.-C., Smail I.},
  J., et~al {Daddi E., Alexander D. M.}, et~al {Daddi E., Bournaud F.}, et~al
  {Daddi E., Cimatti A.}, et~al {Daddi E., Dickinson M.}, et~al {Danielson A.
  L. R., Swinbank A. M.}, {Dannerbauer H.}, G., et~al {Dav{\'{e}} R., Finlator
  K.}, et~al {De Breuck C., Williams R. J.}, {Dekel A.}, D., et~al {Dekel A.,
  Zolotov A.}, et~al {Eales S., Lilly S.}, {Elmegreen B. G.}, M., M., D.,
  {Elmegreen B. G., Elmegreen D. M.}, J., {Elmegreen D. M.}, M., et~al {Engel
  H., Tacconi L. J.}, et~al {F{\"{o}}rster Schreiber N. M., Shapley A. E.},
  et~al {F{\"{o}}rster Schreiber N. M., Genzel R.}, et~al {Genzel R., Newman
  S.}, et~al {Genzel R., Tacconi L. J.}, et~al {Genzel R., Tacconi L. J.},
  et~al {Genzel R., Burkert A.}, et~al {Gilli R., Norman C.}, et~al {Guo Y.,
  Ferguson H. C.}, {Guo Y., Giavalisco M., Ferguson H. C.}, M., F., {\"{O}}zel,
  et~al {Hatsukade B., Tamura Y.}, et~al {Hayward C. C., Jonsson P.}, et~al
  {Hayward C. C., Kere{\v{s}} D.}, et~al {Hodge J. A., Carilli C. L.}, et~al
  {Hodge J. A., Carilli C. L.}, et~al {Hodge J. A., Karim A.}, et~al {Hodge J.
  A., Riechers D.}, et~al {Hwang H. S., Elbaz D.}, et~al {Ikarashi S., Ivison
  R. J.}, et~al {Ivison R. J., Papadopoulos P. P.}, et~al {Ivison R. J., Smail
  I.}, et~al {Karim A., Swinbank A. M.}, {Kere{\v{s}} D., Katz N., Weinberg
  Dav{\'{e}} H. and Dav{\'{e}} R., Fardal M.}, R., H., {Kere{\v{s}} D., Katz
  N., Weinberg Dav{\'{e}} H. and Dav{\'{e}} R., Fardal M.}, R., H.,
  {Kere{\v{s}} D., Katz N., Weinberg Dav{\'{e}} H. and Dav{\'{e}} R., Fardal
  M.}, R., H., et~al {Kov{\'{a}}cs A., Omont A.}, et~al {Lutz D., Berta S.},
  et~al {Mandelker N., Dekel A.}, et~al {Mayer L., Tamburello V.}, et~al
  {Miettinen O., Novak M.}, et~al {Narayanan D., Hayward C. C.}, et~al
  {Narayanan D., Turk M.}, M., et~al {Oklopcic A., Hopkins P. F.}, et~al {Oteo
  I., Ivison R. J.}, et~al {Rich J. W., de Blok W. J. G.}, et~al {Sakamoto K.,
  Wang J.}, F., I., et~al {Scoville N., Murchikova L.}, et~al {Shapiro K. L.,
  Genzel R.}, et~al {Simpson J. M., Swinbank A. M., Smail I.}, et~al {Simpson
  J. M., Swinbank A. M., Smail I.}, et~al {Simpson J. M., Swinbank A. M., Smail
  I.}, et~al {Smol{\v{c}}i{\'{c}} V., Aravena M.}, {Solomon P. M., Rivolo A.
  R.}, A., A., P., et~al {Spergel D. N., Verde L.}, et~al {Spergel D. N., Bean
  R.}, et~al {Steidel C. C., Shapley A. E.}, et~al {Swinbank A. M., Chapman S.
  C.}, et~al {Swinbank A. M., Papadopoulos P. P.}, et~al {Swinbank A. M.,
  Simpson J. M., Smail I., et al Chapman S. C. et al Longmore Sobral et al D.},
  et~al {Swinbank A. M., Simpson J. M., Smail I., et al Chapman S. C. et al
  Longmore Sobral et al D.}, et~al {Swinbank A. M., Simpson J. M., Smail I., et
  al Chapman S. C. et al Longmore Sobral et al D.}, et~al {Swinbank A. M.,
  Simpson J. M., Smail I., et al Chapman S. C. et al Longmore Sobral et al D.},
  et~al {Tacconi L. J., Genzel R.}, et~al {Tacconi L. J., Neri R.}, et~al {Wang
  S. X., Brandt W. N.}, {Wang W.-H., Cowie L. L.}, P., et~al {Wei{\ss} A.,
  Kovacs A.}, Z., Ma, \& et~al {Younger J. D., Fazio G. G.}}]{Hodge2016}
Hodge, J.~A., Swinbank, A.~M., Simpson, J.~M., {et~al.} 2016, \apj, 833, 103,
  \dodoi{10.3847/1538-4357/833/1/103}

\bibitem[{Hodge {et~al.}(2025)Hodge, da~Cunha, Kendrew, Li, Smail, Westoby,
  Nayak, Swinbank, Chen, Walter, van~der Werf, Cracraft, Battisti, Brandt,
  {Calistro Rivera}, Chapman, Cox, Dannerbauer, Decarli, {Frias Castillo},
  Greve, Knudsen, Leslie, Menten, Rybak, Schinnerer, Wardlow, \&
  Weiss}]{Hodge2025}
Hodge, J.~A., da~Cunha, E., Kendrew, S., {et~al.} 2025, \apj, 978, 165,
  \dodoi{10.3847/1538-4357/ad9a52}

\bibitem[{Huang {et~al.}(2023)Huang, Li, Cheng, Hou, Yan, Willner, Dai, Zheng,
  Pan, Rigopoulou, Wang, Li, Liang, Esamdin, \& Fazio}]{Huang2023}
Huang, J.~S., Li, Z.-J., Cheng, C., {et~al.} 2023.
\newblock \doarXiv{2304.01378}

\bibitem[{Hunter(2007)}]{hunter2007}
Hunter, J.~D. 2007, Comput. Sci. Eng., 9, 90, \dodoi{10.1109/MCSE.2007.55}

\bibitem[{Ikeda {et~al.}(2022)Ikeda, Tadaki, Iono, Kodama, Chan, Hatsukade,
  Hayashi, Izumi, Kohno, Koyama, Shimakawa, Suzuki, Tamura, \&
  Tanaka}]{Ikeda2022}
Ikeda, R., Tadaki, K.-i., Iono, D., {et~al.} 2022, \apj, 933, 11,
  \dodoi{10.3847/1538-4357/ac6cdc}

\bibitem[{Kolupuri {et~al.}(2025)Kolupuri, Decarli, Neri, Cox, Ferkinhoff,
  Bertoldi, Weiss, Venemans, Riechers, {Paolo Farina}, \&
  Walter}]{Kolupuri2025}
Kolupuri, S., Decarli, R., Neri, R., {et~al.} 2025, \aap, 695, A201,
  \dodoi{10.1051/0004-6361/202452374}

\bibitem[{Krumholz \& McKee(2005)}]{Krumholz2005}
Krumholz, M.~R., \& McKee, C.~F. 2005, \apj, 630, 250, \dodoi{10.1086/431734}

\bibitem[{Lacy {et~al.}(2017)Lacy, Sneden, Kim, \& Jaffe}]{Lacy2017}
Lacy, J.~H., Sneden, C., Kim, H., \& Jaffe, D.~T. 2017, \apj, 838, 66,
  \dodoi{10.3847/1538-4357/aa6247}

\bibitem[{Leroy {et~al.}(2017)Leroy, Usero, Schruba, Bigiel, Kruijssen, Kepley,
  Blanc, Bolatto, Cormier, Gallagher, Hughes, Jim{\'{e}}nez-Donaire,
  Rosolowsky, \& Schinnerer}]{Leroy2017}
Leroy, A.~K., Usero, A., Schruba, A., {et~al.} 2017, \apj, 835, 217,
  \dodoi{10.3847/1538-4357/835/2/217}

\bibitem[{Li \& Draine(2001)}]{Li2001}
Li, A., \& Draine, B.~T. 2001, \apj, 554, 778, \dodoi{10.1086/323147}

\bibitem[{Magdis {et~al.}(2011)Magdis, Daddi, Elbaz, Sargent, Dickinson,
  Dannerbauer, Aussel, Walter, Hwang, Charmandaris, Hodge, Riechers,
  Rigopoulou, Carilli, Pannella, Mullaney, Leiton, \& Scott}]{Magdis2011}
Magdis, G.~E., Daddi, E., Elbaz, D., {et~al.} 2011, \apj, 740, L15,
  \dodoi{10.1088/2041-8205/740/1/L15}

\bibitem[{Maseda {et~al.}(2024)Maseda, de~Graaff, Franx, Rix, Carniani,
  Laseter, Dudzevi{\v{c}}iūtė, Rawle, Parlanti, Arribas, Bunker, Cameron,
  Charlot, Curti, D'Eugenio, Jones, Kumari, Maiolino, {\"{U}}bler, Saxena,
  Smit, Willott, \& Witstok}]{Maseda2024}
Maseda, M.~V., de~Graaff, A., Franx, M., {et~al.} 2024, \aap, 689, A73,
  \dodoi{10.1051/0004-6361/202449914}

\bibitem[{Padoan \& Nordlund(2002)}]{Padoan2002}
Padoan, P., \& Nordlund, A. 2002, \apj, 576, 870, \dodoi{10.1086/341790}

\bibitem[{Patra {et~al.}(2025)Patra, Evans, Kim, Heyer, Giannetti, Elia, Jose,
  Kauffmann, Samal, Karska, Das, Lee, \& Park}]{Patra2025}
Patra, S., Evans, N.~J., Kim, K.-T., {et~al.} 2025, \apj, 983, 133,
  \dodoi{10.3847/1538-4357/adbf8d}

\bibitem[{Peng {et~al.}(2002)Peng, Ho, Impey, \& Rix}]{Peng2002}
Peng, C.~Y., Ho, L.~C., Impey, C.~D., \& Rix, H.-W. 2002, \aj, 124, 266,
  \dodoi{10.1086/340952}

\bibitem[{Perera {et~al.}(2008)Perera, Chapin, Austermann, Scott, Wilson,
  Halpern, Pope, Scott, Yun, Lowenthal, Morrison, Aretxaga, Bock, Coppin,
  Crowe, Frey, Hughes, Kang, Kim, \& Mauskopf}]{Perera2008}
Perera, T.~A., Chapin, E.~L., Austermann, J.~E., {et~al.} 2008, \mnras, 391,
  1227, \dodoi{10.1111/j.1365-2966.2008.13902.x}

\bibitem[{Perez \& Granger(2007)}]{perez2007}
Perez, F., \& Granger, B.~E. 2007, Comput. Sci. Eng., 9, 21,
  \dodoi{10.1109/MCSE.2007.53}

\bibitem[{P{\'{e}}rez-Gonz{\'{a}}lez {et~al.}(2023)P{\'{e}}rez-Gonz{\'{a}}lez,
  Costantin, Langeroodi, Rinaldi, Annunziatella, Ilbert, Colina,
  N{\o}rgaard-Nielsen, Greve, {\"{O}}stlin, Wright, Alonso-Herrero,
  {\'{A}}lvarez-M{\'{a}}rquez, Caputi, Eckart, {Le F{\`{e}}vre}, Labiano,
  Garc{\'{i}}a-Mar{\'{i}}n, Hjorth, Kendrew, Pye, Tikkanen, van~der Werf,
  Walter, Ward, Bik, Boogaard, Bosman, G{\'{o}}mez, Gillman, Iani, Jermann,
  Melinder, Meyer, Moutard, van Dishoek, Henning, Lagage, Guedel, Peissker,
  Ray, Vandenbussche, Garc{\'{i}}a-Argum{\'{a}}nez, \& {Mar{\'{i}}a
  M{\'{e}}rida}}]{Perez-Gonzalez2023}
P{\'{e}}rez-Gonz{\'{a}}lez, P.~G., Costantin, L., Langeroodi, D., {et~al.}
  2023, \apjl, 951, L1, \dodoi{10.3847/2041-8213/acd9d0}

\bibitem[{{Planck Collaboration} {et~al.}(2020){Planck Collaboration}, Aghanim,
  Akrami, Ashdown, Aumont, Baccigalupi, Ballardini, Banday, Barreiro, Bartolo,
  Basak, Battye, Benabed, Bernard, Bersanelli, Bielewicz, Bock, Bond, Borrill,
  Bouchet, Boulanger, Bucher, Burigana, Butler, Calabrese, Cardoso, Carron,
  Challinor, Chiang, Chluba, Colombo, Combet, Contreras, Crill, Cuttaia,
  de~Bernardis, de~Zotti, Delabrouille, Delouis, {Di Valentino}, Diego,
  Dor{\'{e}}, Douspis, Ducout, Dupac, Dusini, Efstathiou, Elsner, En{\ss}lin,
  Eriksen, Fantaye, Farhang, Fergusson, Fernandez-Cobos, Finelli, Forastieri,
  Frailis, Fraisse, Franceschi, Frolov, Galeotta, Galli, Ganga,
  G{\'{e}}nova-Santos, Gerbino, Ghosh, Gonz{\'{a}}lez-Nuevo, G{\'{o}}rski,
  Gratton, Gruppuso, Gudmundsson, Hamann, Handley, Hansen, Herranz,
  Hildebrandt, Hivon, Huang, Jaffe, Jones, Karakci, Keih{\"{a}}nen, Keskitalo,
  Kiiveri, Kim, Kisner, Knox, Krachmalnicoff, Kunz, Kurki-Suonio, Lagache,
  Lamarre, Lasenby, Lattanzi, Lawrence, {Le Jeune}, Lemos, Lesgourgues,
  Levrier, Lewis, Liguori, Lilje, Lilley, Lindholm, L{\'{o}}pez-Caniego, Lubin,
  Ma, Mac{\'{i}}as-P{\'{e}}rez, Maggio, Maino, Mandolesi, Mangilli,
  Marcos-Caballero, Maris, Martin, Martinelli, Mart{\'{i}}nez-Gonz{\'{a}}lez,
  Matarrese, Mauri, McEwen, Meinhold, Melchiorri, Mennella, Migliaccio, Millea,
  Mitra, Miville-Desch{\^{e}}nes, Molinari, Montier, Morgante, Moss, Natoli,
  N{\o}rgaard-Nielsen, Pagano, Paoletti, Partridge, Patanchon, Peiris,
  Perrotta, Pettorino, Piacentini, Polastri, Polenta, Puget, Rachen, Reinecke,
  Remazeilles, Renzi, Rocha, Rosset, Roudier, Rubi{\~{n}}o-Mart{\'{i}}n,
  Ruiz-Granados, Salvati, Sandri, Savelainen, Scott, Shellard, Sirignano,
  Sirri, Spencer, Sunyaev, Suur-Uski, Tauber, Tavagnacco, Tenti, Toffolatti,
  Tomasi, Trombetti, Valenziano, Valiviita, {Van Tent}, Vibert, Vielva, Villa,
  Vittorio, Wandelt, Wehus, White, White, Zacchei, \&
  Zonca}]{PlanckCollaboration2018a}
{Planck Collaboration}, Aghanim, N., Akrami, Y., {et~al.} 2020, \aap, 641, A6,
  \dodoi{10.1051/0004-6361/201833910}

\bibitem[{Pope {et~al.}(2005)Pope, Borys, Scott, Conselice, Dickinson, \&
  Mobasher}]{Pope2005}
Pope, A., Borys, C., Scott, D., {et~al.} 2005, \mnras, 358, 149,
  \dodoi{10.1111/j.1365-2966.2005.08759.x}

\bibitem[{Pope {et~al.}(2006)Pope, Scott, Dickinson, Chary, Morrison, Borys,
  Sajina, Alexander, Daddi, Frayer, MacDonald, \& Stern}]{Pope2006}
Pope, A., Scott, D., Dickinson, M., {et~al.} 2006, \mnras, 370, 1185,
  \dodoi{10.1111/j.1365-2966.2006.10575.x}

\bibitem[{Popping {et~al.}(2022)Popping, Pillepich, {Calistro Rivera}, Schulz,
  Hernquist, Kaasinen, Marinacci, Nelson, \& Vogelsberger}]{Popping2022}
Popping, G., Pillepich, A., {Calistro Rivera}, G., {et~al.} 2022, \mnras, 510,
  3321, \dodoi{10.1093/mnras/stab3312}

\bibitem[{R{\'{e}}my-Ruyer {et~al.}(2014)R{\'{e}}my-Ruyer, Madden, Galliano,
  Galametz, Takeuchi, Asano, Zhukovska, Lebouteiller, Cormier, Jones, Bocchio,
  Baes, Bendo, Boquien, Boselli, DeLooze, Doublier-Pritchard, Hughes,
  Karczewski, \& Spinoglio}]{Remy-Ruyer2014}
R{\'{e}}my-Ruyer, A., Madden, S.~C., Galliano, F., {et~al.} 2014, \aap, 563,
  A31, \dodoi{10.1051/0004-6361/201322803}

\bibitem[{Riechers {et~al.}(2014)Riechers, Pope, Daddi, Armus, Carilli, Walter,
  Hodge, Chary, Morrison, Dickinson, Dannerbauer, \& Elbaz}]{Riechers2014}
Riechers, D.~A., Pope, A., Daddi, E., {et~al.} 2014, \apj, 786,
  \dodoi{10.1088/0004-637X/786/1/31}

\bibitem[{Rizzo {et~al.}(2024)Rizzo, Bacchini, Kohandel, {Di Mascolo},
  Fraternali, Roman-Oliveira, Zanella, Popping, Valentino, Magdis, \&
  Whitaker}]{Rizzo2024}
Rizzo, F., Bacchini, C., Kohandel, M., {et~al.} 2024, \aap, 689, A273,
  \dodoi{10.1051/0004-6361/202450455}

\bibitem[{Sandstrom {et~al.}(2013)Sandstrom, Leroy, Walter, Bolatto, Croxall,
  Draine, Wilson, Wolfire, Calzetti, Kennicutt, Aniano, {Donovan Meyer}, Usero,
  Bigiel, Brinks, de~Blok, Crocker, Dale, Engelbracht, Galametz, Groves, Hunt,
  Koda, Kreckel, Linz, Meidt, Pellegrini, Rix, Roussel, Schinnerer, Schruba,
  Schuster, Skibba, van~der Laan, Appleton, Armus, Brandl, Gordon, Hinz,
  Krause, Montiel, Sauvage, Schmiedeke, Smith, \& Vigroux}]{Sandstrom2013}
Sandstrom, K.~M., Leroy, A.~K., Walter, F., {et~al.} 2013, \apj, 777, 5,
  \dodoi{10.1088/0004-637X/777/1/5}

\bibitem[{Scoville {et~al.}(2017)Scoville, Lee, Bout, Diaz-Santos, Sanders,
  Darvish, Bongiorno, Casey, Murchikova, Koda, Capak, Vlahakis, Ilbert, Sheth,
  Morokuma-Matsui, Ivison, Aussel, Laigle, McCracken, Armus, Pope, Toft, \&
  Masters}]{Scoville2017}
Scoville, N., Lee, N., Bout, P.~V., {et~al.} 2017, \apj, 837, 150,
  \dodoi{10.3847/1538-4357/aa61a0}

\bibitem[{S{\'{e}}rsic(1963)}]{Sersic1963}
S{\'{e}}rsic, J. 1963, Bol. la Asoc. Argentina Astron. La Plata Argentina, 6,
  41

\bibitem[{S{\'{e}}rsic(1968)}]{Sersic1968}
S{\'{e}}rsic, J.~L. 1968, {Atlas de Galaxias Australes}

\bibitem[{Simpson {et~al.}(2015)Simpson, Smail, Swinbank, Almaini, Blain,
  Bremer, Chapman, Chen, Conselice, Coppin, Danielson, Dunlop, Edge, Farrah,
  Geach, Hartley, Ivison, Karim, Lani, Ma, Meijerink, Micha{\l}owski, Mortlock,
  Scott, Simpson, Spaans, Thomson, {Van Kampen}, \& {Van Der
  Werf}}]{Simpson2015a}
Simpson, J.~M., Smail, I., Swinbank, A.~M., {et~al.} 2015, \apj, 799,
  \dodoi{10.1088/0004-637X/799/1/81}

\bibitem[{Sofia {et~al.}(2004)Sofia, Lauroesch, Meyer, \&
  Cartledge}]{Sofia2004}
Sofia, U.~J., Lauroesch, J.~T., Meyer, D.~M., \& Cartledge, S. I.~B. 2004,
  \apj, 605, 272, \dodoi{10.1086/382592}

\bibitem[{Sokal(1996)}]{Sokal1996}
Sokal, A.~D. 1996.
\newblock \url{https://api.semanticscholar.org/CorpusID:14817657}

\bibitem[{Tadaki {et~al.}(2023)Tadaki, Kodama, Koyama, Suzuki, Mitsuhashi, \&
  Ikeda}]{Tadaki2023}
Tadaki, K.-i., Kodama, T., Koyama, Y., {et~al.} 2023, \apjl, 957, L15,
  \dodoi{10.3847/2041-8213/ad03f2}

\bibitem[{Tadaki {et~al.}(2020)Tadaki, Belli, Burkert, Dekel, {F{\"{o}}rster
  Schreiber}, Genzel, Hayashi, Herrera-Camus, Kodama, Kohno, Koyama, Lee, Lutz,
  Mowla, Nelson, Renzini, Suzuki, Tacconi, {\"{U}}bler, Wisnioski, \&
  Wuyts}]{Tadaki2020}
Tadaki, K.-i., Belli, S., Burkert, A., {et~al.} 2020, \apj, 901, 74,
  \dodoi{10.3847/1538-4357/abaf4a}

\bibitem[{Tan {et~al.}(2014)Tan, Daddi, Magdis, Pannella, Sargent, Riechers,
  B{\'{e}}thermin, Bournaud, Carilli, da~Cunha, Dannerbauer, Dickinson, Elbaz,
  Gao, Hodge, Owen, \& Walter}]{Tan2014}
Tan, Q., Daddi, E., Magdis, G., {et~al.} 2014, \aap, 569, A98,
  \dodoi{10.1051/0004-6361/201423905}

\bibitem[{Taylor(2005)}]{taylor2005}
Taylor, M.~B. 2005, Astron. Data Anal. Softw. Syst. XIV, 347, 29.
\newblock \url{https://ui.adsabs.harvard.edu/abs/2005ASPC..347...29T}

\bibitem[{{The Astropy Collaboration} {et~al.}(2022){The Astropy
  Collaboration}, Price-Whelan, Lim, Earl, Starkman, Bradley, Shupe, Patil,
  Corrales, Brasseur, N{\"{o}}the, Donath, Tollerud, Morris, Ginsburg, Vaher,
  Weaver, Tocknell, Jamieson, van Kerkwijk, Robitaille, Merry, Bachetti,
  G{\"{u}}nther, Aldcroft, Alvarado-Montes, Archibald, B{\'{o}}di, Bapat,
  Barentsen, Baz{\'{a}}n, Biswas, Boquien, Burke, Cara, Cara, Conroy, Conseil,
  Craig, Cross, Cruz, D'Eugenio, Dencheva, Devillepoix, Dietrich, Eigenbrot,
  Erben, Ferreira, Foreman-Mackey, Fox, Freij, Garg, Geda, Glattly,
  Gondhalekar, Gordon, Grant, Greenfield, Groener, Guest, Gurovich, Handberg,
  Hart, Hatfield-Dodds, Homeier, Hosseinzadeh, Jenness, Jones, Joseph,
  Kalmbach, Karamehmetoglu, Ka{\l}uszy{\'{n}}ski, Kelley, Kern, Kerzendorf,
  Koch, Kulumani, Lee, Ly, Ma, MacBride, Maljaars, Muna, Murphy, Norman,
  O'Steen, Oman, Pacifici, Pascual, Pascual-Granado, Patil, Perren, Pickering,
  Rastogi, Roulston, Ryan, Rykoff, Sabater, Sakurikar, Salgado, Sanghi,
  Saunders, Savchenko, Schwardt, Seifert-Eckert, Shih, Jain, Shukla, Sick,
  Simpson, Singanamalla, Singer, Singhal, Sinha, Sipőcz, Spitler, Stansby,
  Streicher, {\v{S}}umak, Swinbank, Taranu, Tewary, Tremblay, de~Val-Borro,
  {Van Kooten}, Vasovi{\'{c}}, Verma, {de Miranda Cardoso}, Williams, Wilson,
  Winkel, Wood-Vasey, Xue, Yoachim, Zhang, \&
  Zonca}]{TheAstropyCollaboration2022}
{The Astropy Collaboration}, Price-Whelan, A.~M., Lim, P.~L., {et~al.} 2022,
  \apj, 935, 167, \dodoi{10.3847/1538-4357/ac7c74}

\bibitem[{{\"{U}}bler {et~al.}(2024){\"{U}}bler, D'Eugenio, Perna, Arribas,
  Jones, Bunker, Carniani, Charlot, Maiolino, {Rodr{\'{i}}guez del Pino},
  Willott, B{\"{o}}ker, Cresci, Kumari, Lamperti, Parlanti, Scholtz, \&
  Venturi}]{Ubler2024}
{\"{U}}bler, H., D'Eugenio, F., Perna, M., {et~al.} 2024, \mnras, 533, 4287,
  \dodoi{10.1093/mnras/stae1993}

\bibitem[{van~der Tak {et~al.}(2007)van~der Tak, Black, Sch{\"{o}}ier, Jansen,
  \& van Dishoeck}]{vanderTak2007}
van~der Tak, F. F.~S., Black, J.~H., Sch{\"{o}}ier, F.~L., Jansen, D.~J., \&
  van Dishoeck, E.~F. 2007, \aap, 468, 627, \dodoi{10.1051/0004-6361:20066820}

\bibitem[{van~der Wel {et~al.}(2012)van~der Wel, Bell, H{\"{a}}ussler, McGrath,
  Chang, Guo, McIntosh, Rix, Barden, Cheung, Faber, Ferguson, Galametz, Grogin,
  Hartley, Kartaltepe, Kocevski, Koekemoer, Lotz, Mozena, Peth, \&
  Peng}]{vanderWel2012}
van~der Wel, A., Bell, E.~F., H{\"{a}}ussler, B., {et~al.} 2012, \apjs, 203,
  24, \dodoi{10.1088/0067-0049/203/2/24}

\bibitem[{Virtanen {et~al.}(2020)Virtanen, Gommers, Oliphant, Haberland, Reddy,
  Cournapeau, Burovski, Peterson, Weckesser, Bright, van~der Walt, Brett,
  Wilson, Millman, Mayorov, Nelson, Jones, Kern, Larson, Carey, Polat, Feng,
  Moore, VanderPlas, Laxalde, Perktold, Cimrman, Henriksen, Quintero, Harris,
  Archibald, Ribeiro, Pedregosa, van Mulbregt, \& {SciPy 1.0
  Contributors}}]{scipy2020}
Virtanen, P., Gommers, R., Oliphant, T.~E., {et~al.} 2020, Nat. Methods, 17,
  261, \dodoi{10.1038/s41592-019-0686-2}

\bibitem[{Wei{\ss} {et~al.}(2007)Wei{\ss}, Downes, Neri, Walter, Henkel,
  Wilner, Wagg, \& Wiklind}]{Weiss2007}
Wei{\ss}, A., Downes, D., Neri, R., {et~al.} 2007, Astronomy, 969, 955,
  \dodoi{10.1051/0004-6361}

\end{thebibliography}
\bibliographystyle{aasjournal}

\appendix

\section{Turbulent Non-Equilibrium Radiative Transfer (TUNER) model}
\label{sec:tuner}

We perform self-consistent radiative transfer modeling of the CO and
dust emission using the Turbulent Non-Equilibrium Radiative Transfer
(TUNER) model.  The innovation of the TUNER model is that it uses a
physically motivated (lognormal) gas density distribution to describe
a realistic density distribution with only two parameters, in contrast
to two-, or multi-component models that quickly require many more free
parameters.  An earlier version of the model is described
\cite{Harrington2021} and builds directly on the single-component
models from \cite{Weiss2007}.  The TUNER implementation features a few
key differences compared to the earlier works, notably in the
implementation of the density distribution, as well as a new Bayesian
framework to constrain the posteriors on the parameters.

In brief, we describe the \emph{volume} distribution of \HH\ gas
($dV$) with a lognormal probability distribution
\citep[$dp$;][]{Padoan2002, Krumholz2005}, such that
$dV(\nHH) = dp(n', \dvturb)$, where $n' = \nHH/\overline{\nHH}$ is the
mean density and $\dvturb$ the turbulent velocity width that defines
the width of the distribution via the Mach number
($\mathcal{M} = \dvturb/\sqrt{8\ln2}/c_s$, with $c_s$ the sound
speed).  The density distribution is by default sampled by 50
components on a grid between 10$^{0}$--10$^{10}$\,cm$^{-3}$.  The
volume of each component $i$ is equal to its equivalent path length
under the LVG assumption ($\ell_{\rm equiv} = \dvturb/(dv/dr)$, where
$dv/dr\ [\mathrm{km}\,\mathrm{s}^{-1}]= 3.1 \kvir
\sqrt{\nHH\,[\mathrm{cm}^{-3}]/10^4}$, with virial parameter $\kvir$;
\citealt{Goldsmith2001, Greve2009}) times the fractional source solid
angle, $V_i=\Omega_i \ell_{{\rm equiv},i}$, such that the full
distribution provides the total flux over the full source solid angle
(parameterized by radius $r$ via $\Omega_s = \pi (r/D_A(z))^2$, with
$D_A(z)$ the angular diameter distance at redshift $z$).  The CO and
dust are coupled to the \HH\ distribution via a single [CO/\HH]
abundance and gas-to-dust ratio, \gdr.  The gas and dust temperatures
are not fixed to be isothermal (e.g. \citealt{Leroy2017}) but are
allowed to vary as a power law with density,
$T(\nHH) \propto \nHH^{\gamma_T}$, where typically $\gamma_T \leq 0$,
such that denser gas is colder, and zero implying isothermal gas.  The
\Tkin\ and \Tdust\ temperatures are normalized such that the volume
weighted means correspond to the input temperatures.  The observed
continuum and line emission (in contrast against the background) can
be derived from the equation of radiative transfer:
\begin{align}
  S_{\nu}^{\rm cont, obs} &= \frac{\Omega_{s}}{(1+z)^{3}} \left(B_{\nu}(\Tdust) - B_{\nu}(T_{\rm CMB}(z))\right)\left(1 - e^{-\tau_{\nu}}\right) \label{eq:rt_dust}\\
  S_{\nu_L}^{\rm line, obs} &= \frac{\Omega_{s}}{(1+z)^{3}}\left(B_{\nu_L, {\rm bg}}(\Texc) - I_{\rm \nu_L, bg}(z)\right)\left(1 - e^{-\tau_{\nu_L}}\right), \label{eq:rt_cont}
\end{align}
where $B_{\nu}(T)$ is Planck blackbody function at the relevant
continuum or line frequency and temperature and $\tau_{\nu}$ is the
corresponding optical depth.  The excitation temperatures (\Texc),
optical depths and line fluxes of each density component are solved
using the LVG approximation and a geometrically averaged escape
probability $p_{\rm esc} = (1 - e^{-\tau_{\nu_L}})/\tau_{\nu_{L}}$
\citep[cf.\ RADEX;][]{vanderTak2007} as described in
\citealt{Weiss2007}).  The background radiation field
($I_{\rm \nu_L, bg}(z)$) is equal to half that of the dust (assuming
complete mixing) plus the Cosmic Microwave Background (CMB) at the
redshift of interest.  The model assumes a default temperature floor
of 10\,K above $T_{\rm CMB}$, to avoid hiding material that is
invisible in contrast to the CMB. The dust opacity is described using
the typical power-law assumption
$\kappa_{\nu} = \kappa_0(\nu/\nu_0)^{\beta}$.  The total mass of the
system is equal to the sum of the mass of the individual components,
which is derived as described in \cite{Weiss2007}.

We derive the posteriors using \textsc{emcee}
\citep{Foreman-Mackey2013} that uses the affine invariant sampler from
\cite{Goodman2010}.  We assess whether the sampling has sufficiently
converged, running the chains until the estimates of the integrated
autocorrelation time $\tau$ \citep{Sokal1996} have settled to within
5\% and the length of the chains are at least $50\tau$.  For
efficiency reasons we only store the posterior during optimization and
recompute the full model output for a representative subset of
posterior samples which is used for further analysis.

\section{Tables and corner plots}
\label{sec:corners}

\begin{figure*}[t]
  \centering
  \includegraphics[width=\textwidth]{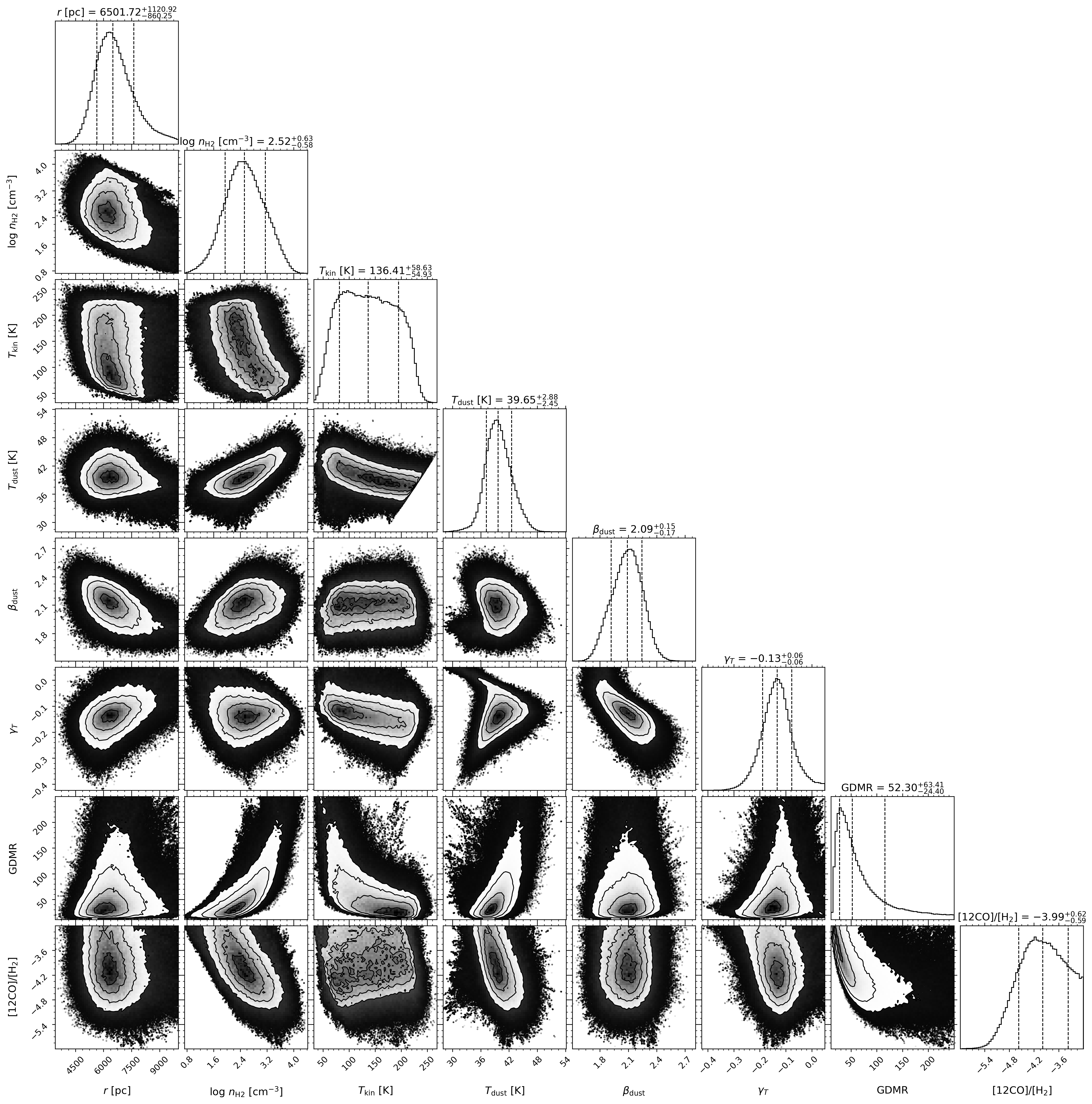}
  \caption{\label{fig:turb_corner} Corner plot
    \citep{Foreman-Mackey2016} of the posterior for the TUNER model
    (\autoref{sec:turb}, see \autoref{fig:turb_fit}).  The top panels
    show the 1D marginalized posteriors with the median and
    16$^{\rm th}$- and 84$^{\rm th}$-percentiles, while the lower
    panels show the 2D marginalized covariances.}
\end{figure*}

\begin{figure*}[t]
  \centering
  \includegraphics[width=\textwidth]{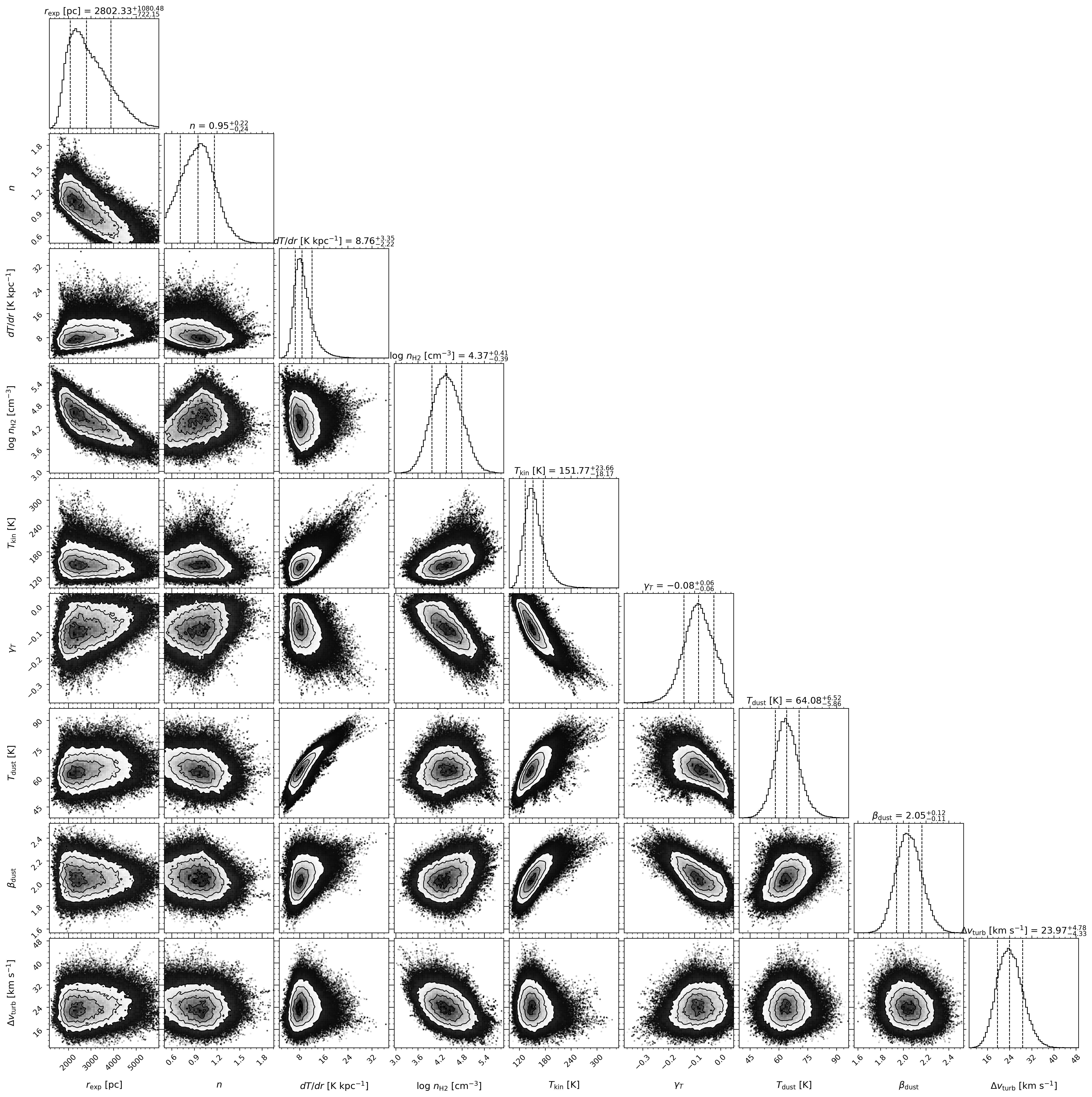}
  \caption{\label{fig:ted_corner} Corner plot
    \citep{Foreman-Mackey2016} of the posterior for the TED model
    (\autoref{sec:ted}, see \autoref{fig:ted_fit}).  The top panels
    show the 1D marginalized posteriors with the median and
    16$^{\rm th}$- and 84$^{\rm th}$ percentiles, while the lower
    panels show the 2D marginalized covariances.  Note the scale
    radius $r_{\rm exp}$ is different from the effective- or
    half-light radius, which is instead computed a
    posteriori.}
\end{figure*}

Continuum and line flux measurements for GN20 are reported in
\autoref{tab:fluxes}.  Corner plots of the posteriors for the models
shown in \autoref{fig:turb_fit} (\autoref{sec:turb}) and
\autoref{fig:ted_fit} (\autoref{sec:ted}) are shown in
\autoref{fig:turb_corner} and \autoref{fig:ted_corner} respectively.
The associated marginalized constraints on the parameters are
tabulated in \autoref{tab:turb} and \autoref{tab:ted}.

\begin{deluxetable}{ccc}[h]
  \tablecaption{GN20 continuum and line measurements.
    \label{tab:fluxes}
  } \tablehead{ \colhead{Parameter} & \colhead{Value} &
    \colhead{Reference} } \startdata
  $S_{\rm 3.3\,mm}$                &   0.23 $\pm$ 0.04  \,mJy              & \citetalias{Tan2014}  \\
  $S_{\rm 3.05\,mm} $             &   0.36 $\pm$ 0.04  \,mJy              & \citetalias{Cortzen2020} \\
  $S_{\rm 2.2\,mm}$                &   0.95 $\pm$ 0.14  \,mJy              & \citetalias{Tan2014}   \\
  $S_{\rm 1.86\,mm}$               &    2.8  $\pm$ 0.13  \,mJy             & \citetalias{Cortzen2020}\\
  $S_{\rm 1.2\,mm}$                  & 8.47 $\pm$ 0.79  \,mJy                & \citetalias{Tan2014}\\
  $S_{\rm 1.1\,mm}$                &   11.5 $\pm$  0.2  \,mJy            &  This work\\
  $S_{\rm 1.1\,mm}$                &   11.5 $\pm$  1.5  \,mJy              &  \citetalias{Perera2008}\\
  $S_{880\,\micron}$          &  18.0 $\pm$ 1.0   \,mJy               &  \citetalias{Hodge2015}\\
  $S_{850\,\micron}$          &  20.3 $\pm$  2.0  \,mJy               &  \citetalias{Pope2006}, \citetalias{Magdis2011}\\
  $S_{500\,\micron}$          &  39.7 $\pm$  6.1  \,mJy               &  \citetalias{Magdis2011}          \\
  $S_{350\,\micron}$          &  41.3 $\pm$  5.2  \,mJy               &  \citetalias{Magdis2011}          \\
  $S_{250\,\micron}$          &  18.66$\pm$  2.70 \,mJy               &  \citetalias{Magdis2011}, \citetalias{Tan2014}\\
  $S_{160\,\micron}^{\dagger}$ &  5.45 $\pm$ 1.02  \,mJy               &  \citetalias{Magdis2011}, \citetalias{Tan2014}\\
  $S_{100\,\micron}^{\dagger}$ &  0.70 $\pm$ 0.24  \,mJy               &  \citetalias{Magdis2011}, \citetalias{Tan2014}\\
  \hline
  $I_{\rm CO(1-0)}$                   &  0.21$\pm$  0.05 \,Jy\,km\,s$^{-1}$ & \citetalias{Carilli2010}\\
  $I_{\rm CO(2-1)}$                  &  0.64 $\pm$  0.16  \,Jy\,km\,s$^{-1}$ & \citetalias{Carilli2010}\\
  $I_{\rm CO(2-1)}^{\ddagger}$      &  0.87 $\pm$ 0.09  \,Jy\,km\,s$^{-1}$ & \citetalias{Carilli2011}\\
  $I_{\rm CO(2-1)}^{\ddagger}$      &  1.0  $\pm$ 0.3    \,Jy\,km\,s$^{-1}$ & \citetalias{Hodge2012}\\
  $I_{\rm CO(4-3)}$                   &  1.68 $\pm$  0.1   \,Jy\,km\,s$^{-1}$ & \citetalias{Tan2014}\\
  $I_{\rm CO(5-4)}$                   &  2.2  $\pm$  0.7   \,Jy\,km\,s$^{-1}$ & \citetalias{Carilli2010}\\
  $I_{\rm CO(6-5)}$                   &  1.8  $\pm$  0.2   \,Jy\,km\,s$^{-1}$ & \citetalias{Carilli2010}\\
  $I_{\rm CO(7-6)}$                   & 2.36 $\pm$ 0.20   \,Jy\,km\,s$^{-1}$ & \citetalias{Cortzen2020}$^{a}$\\
  $I_{\rm \CI(1-0)}^{\ddagger}$                   &  0.70 $\pm$  0.11  \,Jy\,km\,s$^{-1}$ & \citetalias{Cortzen2020}\\
  $I_{\rm \CI(2-1)}^{\ddagger}$ & 1.80 $\pm$ 0.21 \,Jy\,km\,s$^{-1}$ & \citetalias{Cortzen2020}\\
  $I_{\rm \NIIfslb}^{\ddagger}$ & 1.80 $\pm$ 0.21 \,Jy\,km\,s$^{-1}$ &
  \citetalias{Kolupuri2025}
  \enddata

  \tablecomments{$^{\dagger}$Herschel/PACS measurements tracing
    hot/warm dust are excluded in the modeling. $^{\ddagger}$CO(2--1)
    measurements based on higher-resolution data and lines other than
    CO are not used in the modeling, see discussion in \autoref{sec:RT
      modeling}. $^{a}$G. Magdis, private communication.  The CO(7--6)
    and \CI(2--1) measurements supersede the inconsistent upper limits
    from \citep{Casey2009}.  References:
    \citet[][\citetalias{Pope2006}]{Pope2006},
    \citet[][\citetalias{Perera2008}]{Perera2008},
    \citet[][\citetalias{Carilli2010}]{Carilli2010},
    \citet[][\citetalias{Carilli2011}]{Carilli2011},
    \citet[][\citetalias{Magdis2011}]{Magdis2011},
    \citet[][\citetalias{Hodge2012}]{Hodge2012},
    \citet[][\citetalias{Tan2014}]{Tan2014},
    \citet[][\citetalias{Hodge2015}]{Hodge2015},
    \citet[][\citetalias{Cortzen2020}]{Cortzen2020},
    \citet[][\citetalias{Kolupuri2025}]{Kolupuri2025}.}
\end{deluxetable}

\begin{deluxetable}{cc}[ht]  %
  \tablecaption{GN20 TUNER model posterior parameter percentiles. \label{tab:turb} }
  \tablehead{ \colhead{Parameter} & \colhead{Value} }
  \startdata
  $r$ [pc]           & $6511^{+1345}_{-989}$ \\
  log $n_{\rm H2}$ [cm$^{-3}$] & $2.64^{+0.63}_{-0.59}$ \\
  $T_{\rm kin}$ [K]  & $125.8^{+56.1}_{-52.8}$ \\
  $T_{\rm dust}$ [K] & $40.2^{+2.9}_{-2.8}$ \\
  $\beta_{\rm dust}$ & $2.08^{+0.15}_{-0.18}$ \\
  $\gamma_{T}$       & $-0.13^{+0.05}_{-0.05}$ \\
  \gdr               & $57^{+116}_{-28}$ \\
  \COHH $\times 10^{-5}$ & $7.07^{+32.63}_{-4.98}$ \\
  log $n_{\rm H2}^{\rm out}$ [cm$^{-3}$] & $2.82^{+0.53}_{-0.43}$ \\
  $M_{\rm mol}$ [$10^{11}$ M$_{\odot}$] & $3.46^{+5.91}_{-1.81}$ \\
  $M_{\rm dust}$ [$10^{9}$ M$_{\odot}$] & $5.87^{+0.61}_{-0.97}$ \\
  $L_{\rm CO(1-0)}'$[$10^{10}$ K\,km\,s$^{-1}$\,pc$^{2}$] & $9.71^{+1.32}_{-1.36}$ \\
  $\alpha_{\rm CO(1-0)}$ [M$_{\odot}$(K\,km\,s$^{-1}$\,pc$^{2}$)$^{-1}$] & $3.56^{+7.26}_{-1.90}$ \\
  $L_{\rm IR}$ [$10^{13}$ L$_{\odot}$] & $1.56^{+0.12}_{-0.09}$
  \enddata
  \tablecomments{The parameters are described in
    \autoref{sec:RT modeling} and \autoref{sec:tuner}.  The mean
    density of the distribution over the modeled density range is
    $\log n_{\HH}^{\rm out}$ (see \autoref{sec:tuner}).  The infrared
    luminosity is integrated over 8--1000\,\micron.}\vspace{-1cm}
\end{deluxetable}

\begin{deluxetable}{cc}[ht]
  \tablecaption{GN20 TED model posterior parameter percentiles.  \label{tab:ted} }
  \tablehead{ \colhead{Parameter} & \colhead{Value} }
  \startdata
  $r_{\rm exp}$ [pc] & $2914^{+1245}_{-965}$ \\
  $n$                & $0.98^{+0.18}_{-0.30}$ \\
  $dT/dr$ [K kpc$^{-1}$] & $8.42^{+5.11}_{-2.37}$ \\
  log $n_{\rm H2}$ [cm$^{-3}$] & $4.31^{+0.43}_{-0.39}$ \\
  $T_{\rm kin}$ [K]  & $154.6^{+21.8}_{-21.9}$ \\
  $\gamma_{T}$       & $-0.08^{+0.06}_{-0.04}$ \\
  $T_{\rm dust}$ [K] & $62.7^{+9.7}_{-5.6}$ \\
  $\beta_{\rm dust}$ & $2.06^{+0.07}_{-0.12}$ \\
  $\Delta v_{\rm turb}$ [km s$^{-1}$] & $24.9^{+4.3}_{-3.3}$ \\
  log $n_{\rm H2}^{\rm out}$ [cm$^{-3}$] & $3.78^{+0.44}_{-0.35}$ \\
  $M_{\rm mol}$ [$10^{11}$ M$_{\odot}$] & $2.87^{+0.39}_{-0.30}$ \\
  $M_{\rm dust}$ [$10^{9}$ M$_{\odot}$] & $5.74^{+0.79}_{-0.60}$ \\
  $L_{\rm CO(1-0)}'$[$10^{11}$ K\,km\,s$^{-1}$\,pc$^{2}$] & $1.03^{+0.10}_{-0.11}$ \\
  $\alpha_{\rm CO(1-0)}$ [M$_{\odot}$(K\,km\,s$^{-1}$\,pc$^{2}$)$^{-1}$] & $2.76^{+0.52}_{-0.34}$ \\
  $L_{\rm IR}$ [$10^{13}$ L$_{\odot}$] & $1.64^{+0.13}_{-0.14}$ \\
  \enddata
  \tablecomments{Parameter description in \autoref{tab:turb}.}
\end{deluxetable}

\end{document}